\documentclass[reqno,11pt]{article}
\usepackage{amsthm,amsmath,amssymb,graphicx,setspace,verbatim}
\usepackage[authoryear]{natbib}
\usepackage[small]{caption}
\usepackage{wrapfig}
\usepackage{multirow}
\usepackage{subfig}
\usepackage{color}
\usepackage[utf8]{inputenc}

\newcommand{\bzero}{\mbox{\boldmath $0$}}

\newcommand{\bc}{\mbox{\boldmath $c$}}

\newcommand{\br}{\mbox{\boldmath $r$}}

\newcommand{\bu}{\mbox{\boldmath $u$}}
\newcommand{\bv}{\mbox{\boldmath $v$}}

\newcommand{\bx}{\mbox{\boldmath $x$}}

\newcommand{\bz}{\mbox{\boldmath $z$}}

\newcommand{\bI}{\mbox{\boldmath $I$}}

\newcommand{\bQ}{\mbox{\boldmath $Q$}}

\newcommand{\cL}{{\cal L}}

\newcommand{\eps}{\varepsilon}

\newcommand{\talpha}{\tilde{\alpha}}
\newcommand{\tbeta}{\tilde{\beta}}
\newcommand{\tdelta}{\tilde{\delta}}

\newcommand{\tomega}{\tilde{\omega}}
\newcommand{\tgamma}{\tilde{\gamma}}
\newcommand{\tnu}{\tilde{\nu}}
\newcommand{\tvarsigma}{\tilde{\varsigma}}
\newcommand{\ttau}{\tilde{\tau}}

\newcommand{\tmu}{\tilde{\mu}}
\newcommand{\tsigma}{\tilde{\sigma}}

\newcommand{\hTheta}{\hat{\Theta}}

\newcommand{\beps}{\mbox{\boldmath $\varepsilon$}}

\newcommand{\bxi}{\mbox{\boldmath $\xi$}}

\newcommand{\blambda}{\mbox{\boldmath $\lambda$}}
\newcommand{\bmu}{\mbox{\boldmath $\mu$}}

\newcommand{\bpi}{\mbox{\boldmath $\pi$}}

\newcommand{\bomega}{\mbox{\boldmath $\omega$}}

\newcommand{\bSigma}{\mbox{\boldmath $\Sigma$}}

\newcommand{\bGamma}{\mbox{\boldmath $\Gamma$}}

\newcommand{\btomega}{\mbox{\boldmath $\tomega$}}

\newcommand{\E}{\mbox{E}}

\newcommand{\bdm}{\begin{displaymath}}
\newcommand{\edm}{\end{displaymath}}
\newcommand{\beq}{\begin{equation}}
\newcommand{\eeq}{\end{equation}}

\renewcommand{\th}{^{\mbox{\scriptsize th}}}

\newtheorem{prop}{Proposition}

\long\def\symbolfootnote[#1]#2{\begingroup%
\def\thefootnote{\fnsymbol{footnote}}\footnote[#1]{#2}\endgroup}

\begin{document}

\pagenumbering{arabic}
\begin{center}
{\singlespacing
\begin{Large}{\bf
Modeling and Predicting Power Consumption of High Performance Computing Jobs\\}
\vspace{.25in}
\end{Large}

Curtis Storlie$^\dag$, Joe Sexton$^\dag$, Scott Pakin$^\dag$, Michael Lang$^\dag$, \\ Brian Reich$^\ddag$, William Rust$^\dag$ \\[.15in]

$^\dag$ Los Alamos National Laboratory\\
$^\ddag$ North Carolina State University\\

\begin{abstract}
Power is becoming an increasingly important concern for large supercomputing centers.  Due to cost concerns, data centers are becoming increasingly limited in their ability to enhance their power infrastructure to support increased compute power on the machine-room floor.  At Los Alamos National Laboratory it is projected that future-generation supercomputers will be power-limited rather than budget-limited.  That is, it will be less costly to acquire a large number of nodes than it will be to upgrade an existing data-center and machine-room power infrastructure to run that large number of nodes at full power.  In the power-limited systems of the future, machines will in principle be capable of drawing more power than they have available.  Thus, power capping at the node/job level must be used to ensure the total system power draw remains below the available level.  In this paper, we present a statistically grounded framework with which to predict (with uncertainty) how much power a given job will need and use these predictions to provide an optimal node-level power capping strategy.  We model the power drawn by a given job (and subsequently by the entire machine) using hierarchical Bayesian modeling with hidden Markov and Dirichlet process models.  We then demonstrate how this model can be used inside of a power-management scheme to minimize the affect of power capping on user jobs.

\vspace{.075in}
\noindent
\emph{Keywords}: High Performance Computing; Power Consumption; Dirichlet Process; Hierarchical Bayesian Modeling; Hidden Markov; Power Capping.

\vspace{.075in}
\noindent
\emph{Running title}: Predicting Power Consumption of High Performance Computers

\vspace{.075in}
\noindent
\emph{Corresponding Author}: Curtis Storlie, \verb1storlie@lanl.gov1

\end{abstract}
}

%\vspace*{0.10in}
%Date: \today
\end{center}

%%%%%%%%%%%%%%%%%%%%%%%%%%%%%%%%%%%%%%%%%%%%%%%%%%%%%%%%%%%%%%%%%%%%%%%%%%%%%%%%%%%%%%%%%%%

\vspace{-.35in}
\section{Introduction}
\vspace{-.0in}

\vspace{-.15in}
\subsection{Power Concerns for Supercomputers}
\vspace{-.05in}

Power has become an increasingly important concern for large supercomputing centers \citep{Kamil08}.  Due to cost concerns, data centers are nearing their capacity to enhance their power infrastructure to support increased compute power on the machine-room floor \citep{Patki13,Zhang14}.  At Los Alamos National Laboratory it is projected that future-generation machines will be power-limited rather than budget-limited.  That is, it will be less costly to acquire a large number of nodes than it will be to upgrade an existing data-center and machine-room power infrastructure to run that large number of nodes at full power.  This is because the cost of power capacity follows a precipitous step function due to the need for construction work on the building and the installation of power substations, chillers, and other large investments.  That said, it is often the case that there is a substantial amount of \emph{trapped capacity} in existing supercomputing data centers \citep{Pakin13,Zhang14}. That is, more power infrastructure is allocated to existing supercomputers than these machines typically draw.  Trapped capacity is the difference between the infrastructure capacity allocated to a given machine (i.e., supercomputer) and the actual peak demand of that machine. For example, the electrical feeder to a rack of servers is typically sized to be able to feed all of the servers running simultaneously at maximum power draw. However, in normal operation the peak electrical demand of the rack may never exceed even half of the demand that was used to size the feeder.
%Yet the excess capacity is held in reserve “just in case” the worst-case demand were to occur. This reserve excess capacity is referred to as trapped capacity.
Future systems will not have the luxury of this trapped-capacity cushion.  In particular, node-level power capping (i.e., throttling performance to limit power consumption) will need to be used to get the maximum performance out of the available power.  In this work statistical modeling of job power is used to provide an optimal power capping strategy according to a given criterion (e.g., maximize throughput, minimize user inconvenience, etc.).

\begin{figure}[t!]
\vspace{-.1in}
  \begin{center}
\caption{Illustration of trapped-capacity for the Luna supercomputer at LANL.}
\vspace{-.15in}
\includegraphics[width=.84\textwidth]{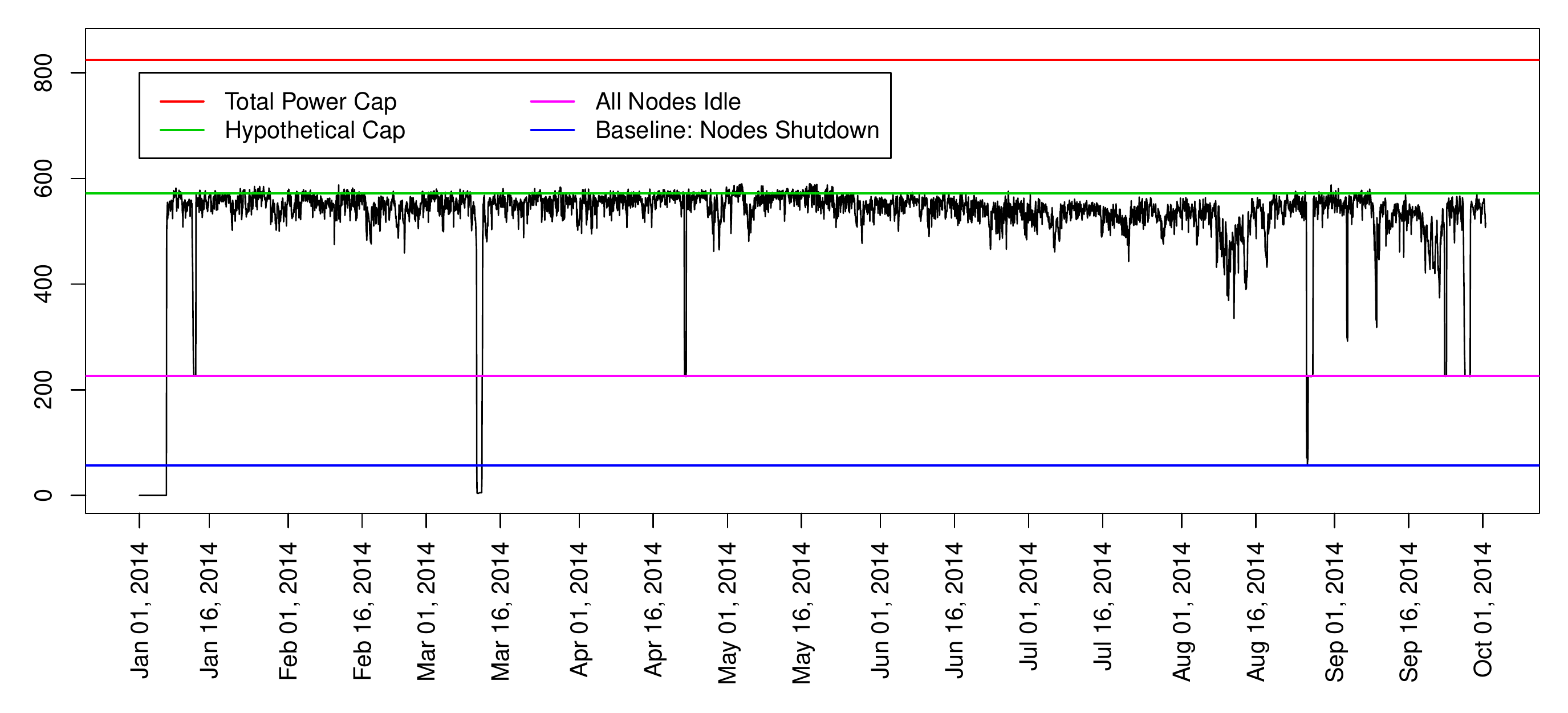}
\label{fig:3machine_pow}
\end{center}
\vspace{-.35in}
\end{figure}

Figure~\ref{fig:3machine_pow}, for example, displays the power drawn from the Luna supercomputer at Los Alamos National Laboratory (LANL) and it is apparent that the peak allocated power could be substantially reduced.
%, or sufficiently more computing hardware (i.e., compute nodes) could be safely addedwithout increasing the power allocated to the machine.
The trapped capacity of Luna is typical of most machines at LANL.  The reason for trapped capacity is that it is difficult to predict what the typical power draw will be for a machine prior to actually running jobs on it.  The power infrastructure for a machine must be developed prior to having any of this information, and it is thus designed to accommodate a conservative estimate of a theoretical peak power draw from the machine.
% This generally results in very conservative situations such as that depicted in Figure~\ref{fig:3machine_pow}.  The cost of the power to run the machine on a day-to-day basis is generally negotiated in a conservative manner as well, i.e., LANL purchases the ability to use far more power than is actually used on a given day.  To date, this conservatism has not caused any problems as the cost of building and running a machine has been dominated by hardware costs.  However, in the foreseeable future, the limiting factor will be the cost of the power needed to run the machine.

Hence, we use as our example for this paper a hypothetical machine \emph{Sol} with the same number of nodes and architecture as Luna, but with a smaller peak power allotment of 575~kW (i.e., the hypothetical cap provided in Figure~\ref{fig:3machine_pow}).  This is intended to mimic the power-limited scenario of the future, where the nodes could, in principle, draw more power collectively than is available to the machine.  Luna is composed of 1540 compute nodes each with two sockets of 8-core, 2.6\,GHz, Intel Xeon E5-2670 processors---a total of 24,640 processors.  While a machine of the future will not necessarily have the same basic architecture as Luna, Luna was chosen as the template for this example because it had most of the data needed for this analysis readily obtainable.  Further, this paper is about a proof of concept using power data, independent of architecture.  Machines of the future, including the newest machine on the horizon at LANL, Trinity, will also have the ability to set a hard cap for the power draw to each node (mainly via CPU throttling) at the expense of performance.  Thus, we also assume that the hypothetical Sol machine has this node-level power-capping capability.  

Suppose the hard cap for each node $i$ was set to $T_i$ such that $\sum_i T_i + B \leq T$~kW, where $B$ is the \emph{baseline} required to power the machine irregardless of whether the nodes are even powered on (e.g., power to network switches, etc.), and $T$ is the peak power available.  The baseline of $B=56.5$~kW and $T=820$~kW for the Luna machine are depicted in Figure~\ref{fig:3machine_pow}.  We assume the Sol machine has the same baseline, but introduce a more stringent hypothetical power cap of $T=575$~kW.  This is to mimic a power-limited future machine that in theory does not have enough power available to run all of its hardware at full throttle.  However, if such a node-level capping constraint above were imposed, then there would not be an issue with going over the power threshold and possibly tripping breakers, damaging nodes, paying exorbitant utility penalties, etc.  This cap can also be adjusted for each node fairly quickly, e.g.,  inside of a minute time frame, which coincides with the frequency of the data observations for this study.  If more frequent observations are available to inform the caps, then the methodology described here can easily be applied to that time scale.
The main question becomes, what is the best way to choose the hard cap for each of the various nodes as jobs are running on them?

A very simplistic approach would be to let each node's cap be $T_i = (T-B)/N$ where $N$ is the number of nodes in the machine.  However, this approach would not allow the nodes that are being worked hard by a compute-intensive job to draw more power than those running closer to idle (e.g.,~because they are blocked on I/O)\@. Instead, we seek to use the power data being collected for each node to make predictions for a short time horizon and use these predictions to determine the best capping strategy for that time frame.  

To accomplish this goal, a stochastic process model is developed for the power drawn by the nodes running the same job.  This model can then be updated as more data become available for a specific job and then used to predict (i.e., produce many realizations of) the power that may be drawn by each node running that job.  These future power realizations can be used to assess the detriment to performance due to a possible hard cap for the nodes running that job.  And then, an optimal, machine wide, node capping strategy can be implemented.

\begin{figure}[t!]
\vspace{-.1in}
\begin{center}
\caption{Cage power over time for three distinct jobs.}
\vspace{-.1in}
\includegraphics[width=.9\textwidth]{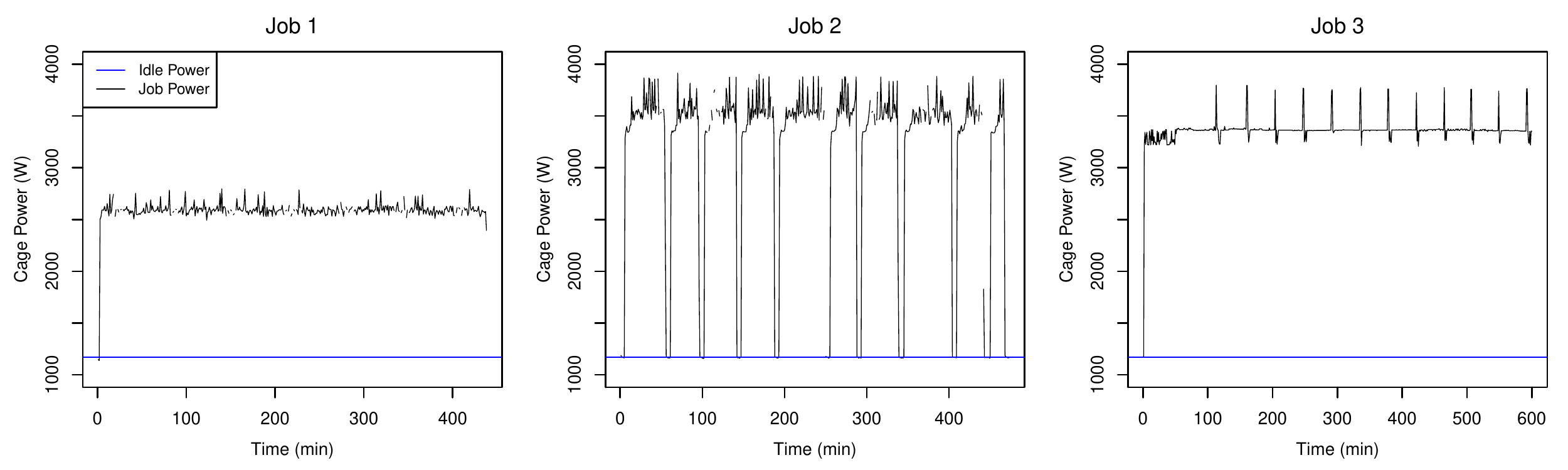}
\label{fig:3_job_pow}
\end{center}
\vspace{-.3in}
\end{figure}

In order to illustrate the concept, we assume that the distribution of jobs (and the power that they draw) on the hypothetical Sol machine is identical to that of Luna.  Luna is \emph{not} currently instrumented to collect power at the node level, which would be ideal.  This is merely an instrumentation/cost issue and node-level power measurement will be available on the new Trinity machine and likely for all future machines.
Luna does, however, currently support power monitoring at the \emph{cage} (10-node) level.  Therefore, all 213 user jobs that took up at least an entire cage during the study period were selected for the analysis presented here.  Figure~\ref{fig:3_job_pow} provides the cage level power draw over time from three such production jobs (i.e., actual scientific compute jobs run on Luna) in the dataset.  %Some summary statistics across all of the 213 jobs are provided in Table~\ref{tab:summary_stats}.
The cage level power draw time series (measured once per minute) for all 213 jobs (454 time series in total since some jobs spanned multiple cages) is available for download at the journal website.

\begin{comment}
\begin{table}[t]
\vspace{-.12in}
  {\small
  \begin{center}
    \caption{Summary statistics across all of the 213 jobs. $^*$For the summary in this table, the number of tasks for a given job was determined from a normal mixture fit with the mclust package in R.}
\label{tab:summary_stats}
\vspace{-.05in}
\begin{tabular}{|l|cccccc|}
  \hline
  Job Variable &  Min & Q1 & Median & Mean & Q3 & Max\\
  \hline
Duration (min) & 10 & 47 & 237 & 265 & 468 & 601\\
Number of Nodes & 10 & 10 & 14 & 23 & 24 & 120\\
Number of Tasks$^*$ & 1 & 2 & 2 & 2.4 & 3 & 5\\
Mean Power (kW) & 1.2 & 3.2 & 3.4 & 3.2 & 3.5 & 3.9\\
Max Power (kW) & 1.2 & 3.5 & 3.7 & 3.6 & 3.8 & 4.3\\
Max - Mean (kW) & 0 & 0.2 & 0.3 & 0.3 & 0.4 & 2.2\\
\hline
\end{tabular}
\end{center}
\vspace{-.28in}
}
\end{table}
\end{comment}

Because of the restriction that these jobs must encompass a cage, this does not constitute a truly random sample of all jobs seen on Luna.  However, most ($> 90$\%) of the nodes being used at a given time on Luna are used by such jobs, and they are typically the more compute-intensive and interesting jobs anyhow.  Thus, for the purpose of ``proving the concept'' of efficient power capping in this work, we make the simplifying assumption that all jobs occurring on Sol come from the same population as jobs spanning a 10-node cage on Luna.  Node-level power measurement is forthcoming for new machines in any case, and an identical approach to that described here would apply directly to node-level data when these data become available.
%Further, this simplifying assumption, if anything, is conservative, since larger jobs typically draw more power per node.  Even still, as demonstrated in Section~\ref{sec:analysis}, the potential benefit of applying the proposed capping approach in practice is substantial.

The total power used by the Sol machine is then equal to the sum of the power draws for the 154 cages, plus the additional baseline level $B$. However, the baseline power draw is nearly constant \citep{Pakin13}.  Thus, the crux of this work is then to provide an accurate probabilistic characterization of the power drawn at the node level (or cage level in this case).
%by a given job.  

\vspace{-.15in}
\subsection{Overview of the Statistical Approach}
\vspace{-.05in}

We propose a sophisticated statistical model for the power profile of a high performance computing (HPC) job.  To accommodate the complex non-Gaussian features illustrated in Figure 2, we use a nonparametric Bayesian model for each job's time series.
A hidden Markov model (HMM) describes the transitions between different regimes (or tasks within the job) and a correlated residual process allows for fluctuation within a task.
%This approach is described in detail in Section~\ref{sec:Bayes_job_power}.
Our approach builds on the emerging literature on hierarchical Dirichlet process hidden Markov (HDP-HM) models \citep{beal:2002, kottas:2009,
Lennox:2010, paisley:2010, Fox:2011}. In these flexible models, the number of potential states in the hidden Markov model is infinite, and the transition probabilities between states are modeled using the hierarchical Dirichlet process prior \citep{Teh:2006}.  Our model is most similar to the sticky-HDP-HM model of \citep{Fox:2011}, who specify a HDP-HM model with added probability on the staying in the same state.

Rather than modeling a single time series, our application requires a joint analysis of multiple jobs. \cite{Fox:2014} also analyze multiple time series using a beta process to share information across series.  In our approach, each job is permitted to have different operating characteristics defined by job-specific parameters (e.g., mean time between state transitions and mean value in each state) modeled as draws from a flexible parent distribution.   This approach facilitates borrowing of strength across jobs to improve prediction for short series, yet allows flexibility to capture complex features as data accrues.  We then demonstrate how this model can be used inside of a power-management scheme to minimize the affect of power capping on user jobs.  Such a scheme will be essential for HPC machines in the power-limited future.  While sophisticated statistical models have been applied to reliability of HPC machines \citep{Storlie12a,Michalak12}, to the best of our knowledge this is the first effort to statistically model the power process of HPC jobs.  This paper also has online supplementary material containing Markov chain Monte Carlo (MCMC) estimation details.

\vspace{-.2in}
\section{The Effect of Power Capping on Job Performance}
\vspace{-.1in}
\label{sec:perf_by_power}

Before diving into a statistical model for the power profile of a job, it is first important to understand how a decrease in available power will affect job performance.  Once the relationship between job performance (i.e., how long the job takes) and available power is understood, the statistical model for power can then be applied to predict the possible degradation to job performance (i.e., increase in run time) due to a given power cap.

\begin{figure}[t]
\vspace{-.11in}
\caption{Derivation of an upper bound on performance degradation: (a) Theoretical bound on Power by CPU frequency along with actual power drawn by the \emph{POP} program. (b) Proportion of full throttle CPU (3.0 GHz) by Proportion of full throttle CPU power (i.e., above and beyond idle power).  (c) Run time by CPU frequency for the POP program.  (d) Theoretical bound on proportion of run time at full throttle CPU that is required to complete a job by proportion of full throttle CPU along with actual for POP.  (e) Relationships depicted in (c) and (d) are combined to produce the upper bound for the proportional run time as a function of the proportion of full power that is available.}
\label{fig:derivation_of_bound}
\vspace{-.05in}
\begin{center}
\begin{minipage}[b][1.6in]{1.3in}
\subfloat[]{\includegraphics[width=1.3in]{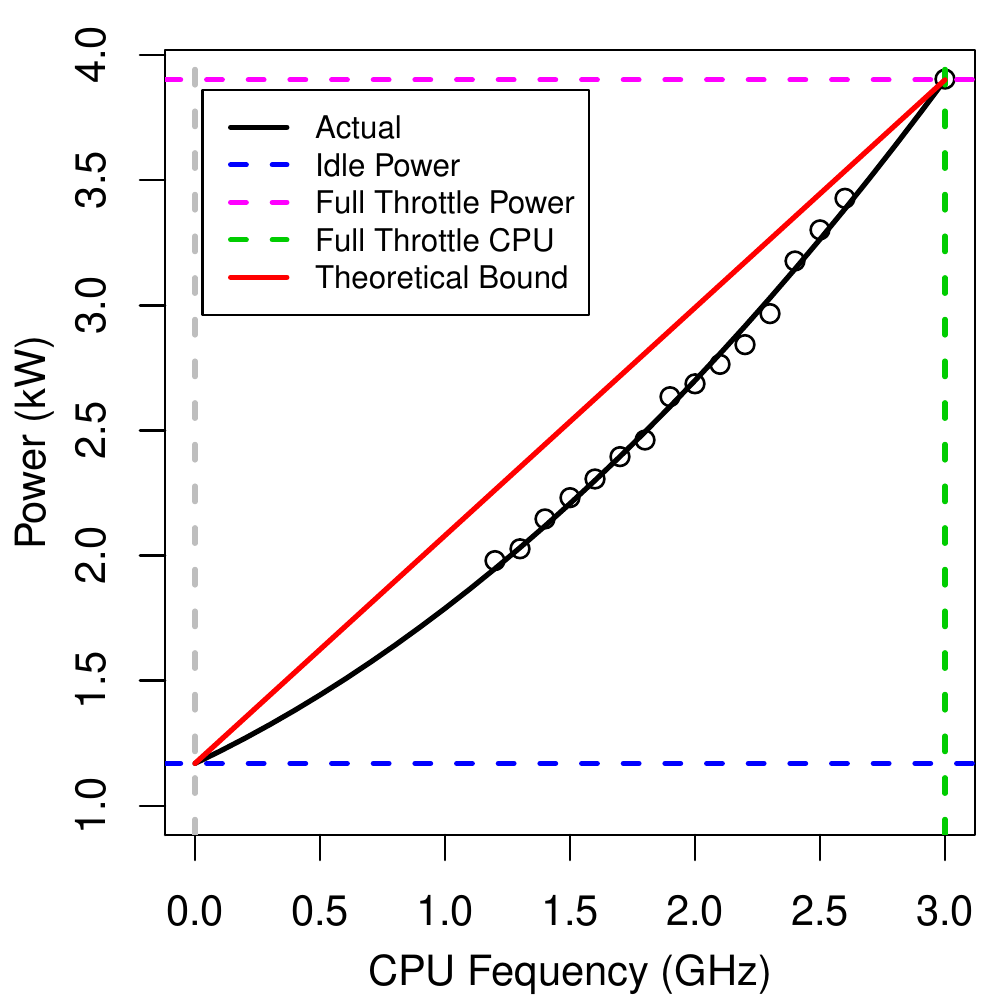}}
\vfill
\setcounter{subfigure}{2}
\subfloat[]{\includegraphics[width=1.3in]{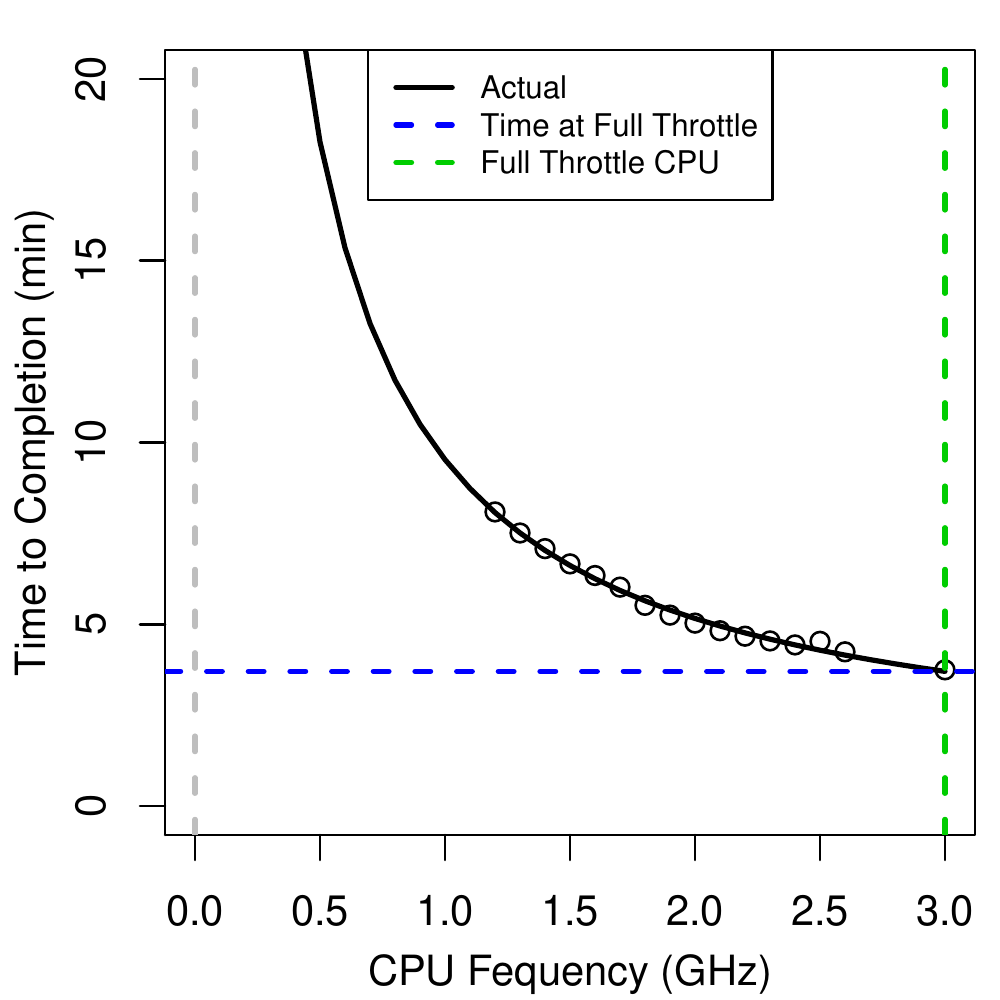}}
\end{minipage}%
\begin{minipage}[b][1.5in]{0.275in}
\raisebox{0.3in}{\includegraphics[width=0.275in]{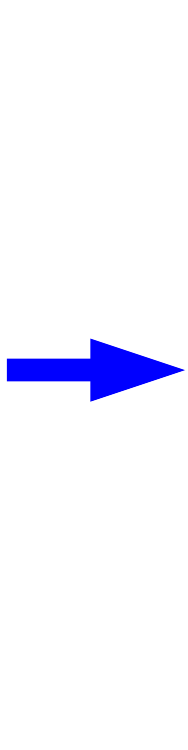}}
\vfill
\raisebox{-1.15in}{\includegraphics[width=0.275in]{line.pdf}}
\end{minipage}%
\begin{minipage}[b][1.6in]{1.3in}
\setcounter{subfigure}{1}
\subfloat[]{\includegraphics[width=1.3in]{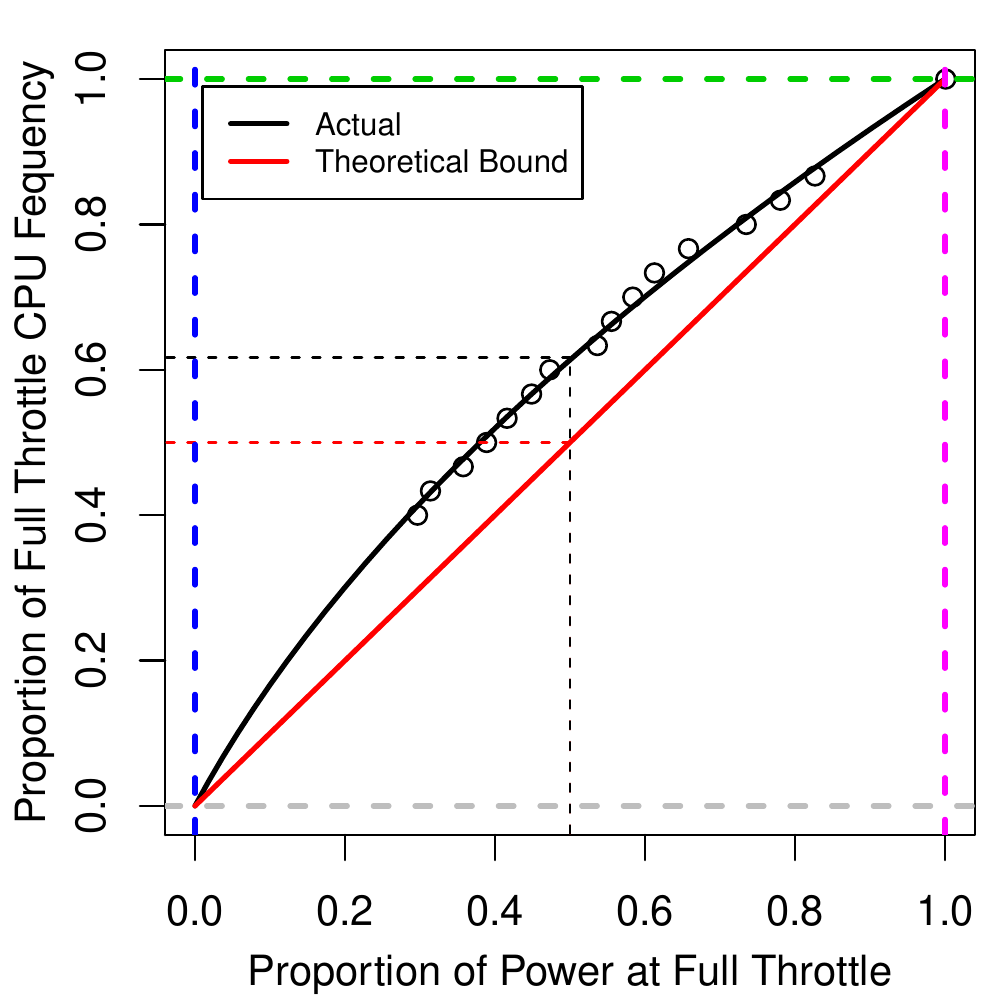}}
\vfill
\setcounter{subfigure}{3}
\subfloat[]{\includegraphics[width=1.3in]{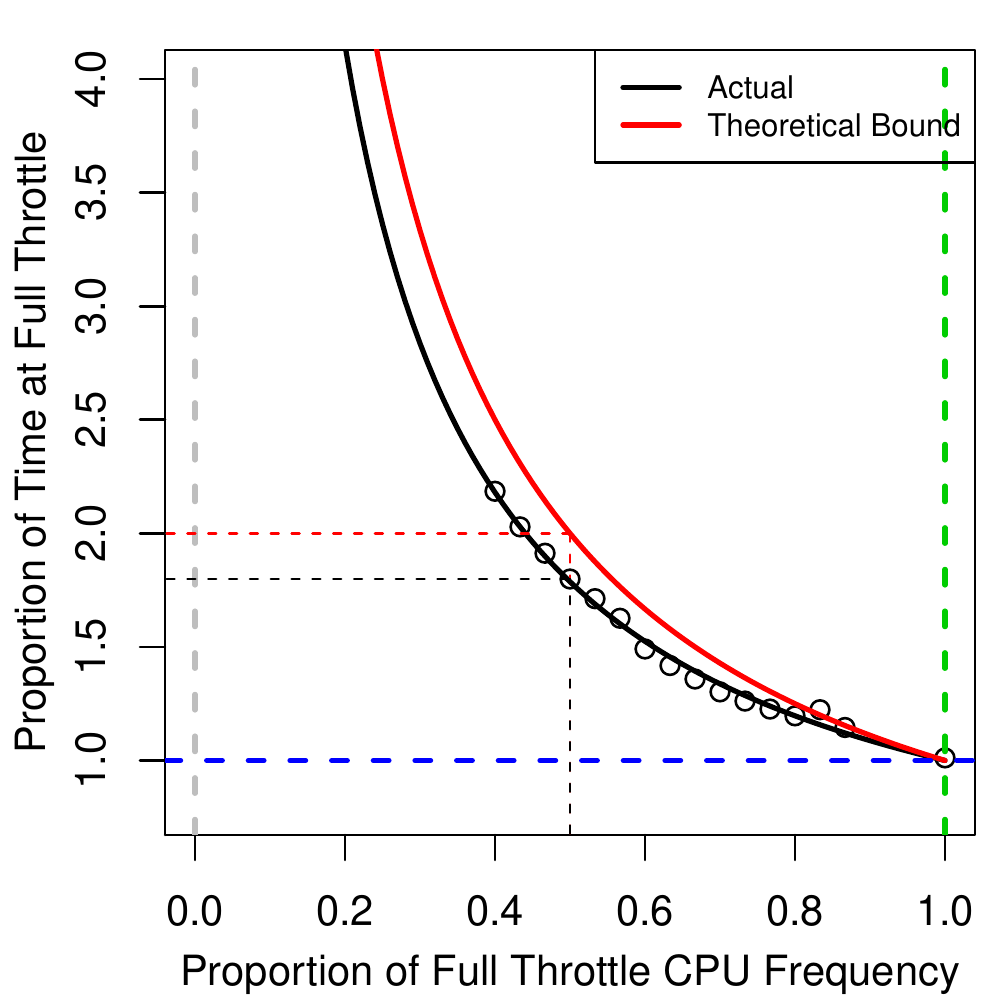}}
\end{minipage}%
\begin{minipage}[t][1.4in]{0.35in}
\raisebox{-.65in}{\includegraphics[width=.35in]{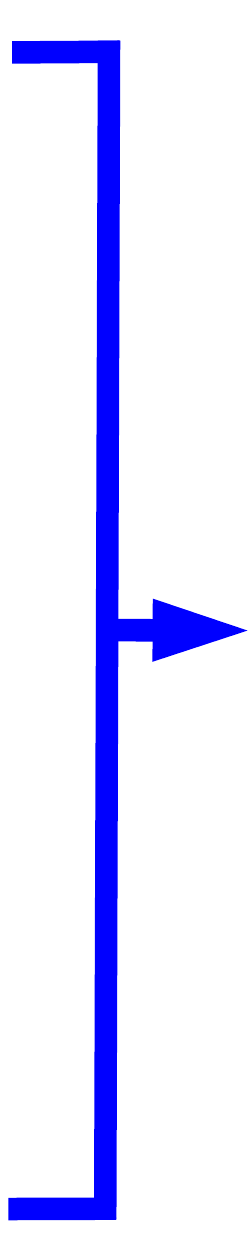}}
\end{minipage}%
\begin{minipage}[t][1.4in]{1.3in}
\raisebox{.85in}{\subfloat[]{\includegraphics[width=1.3in]{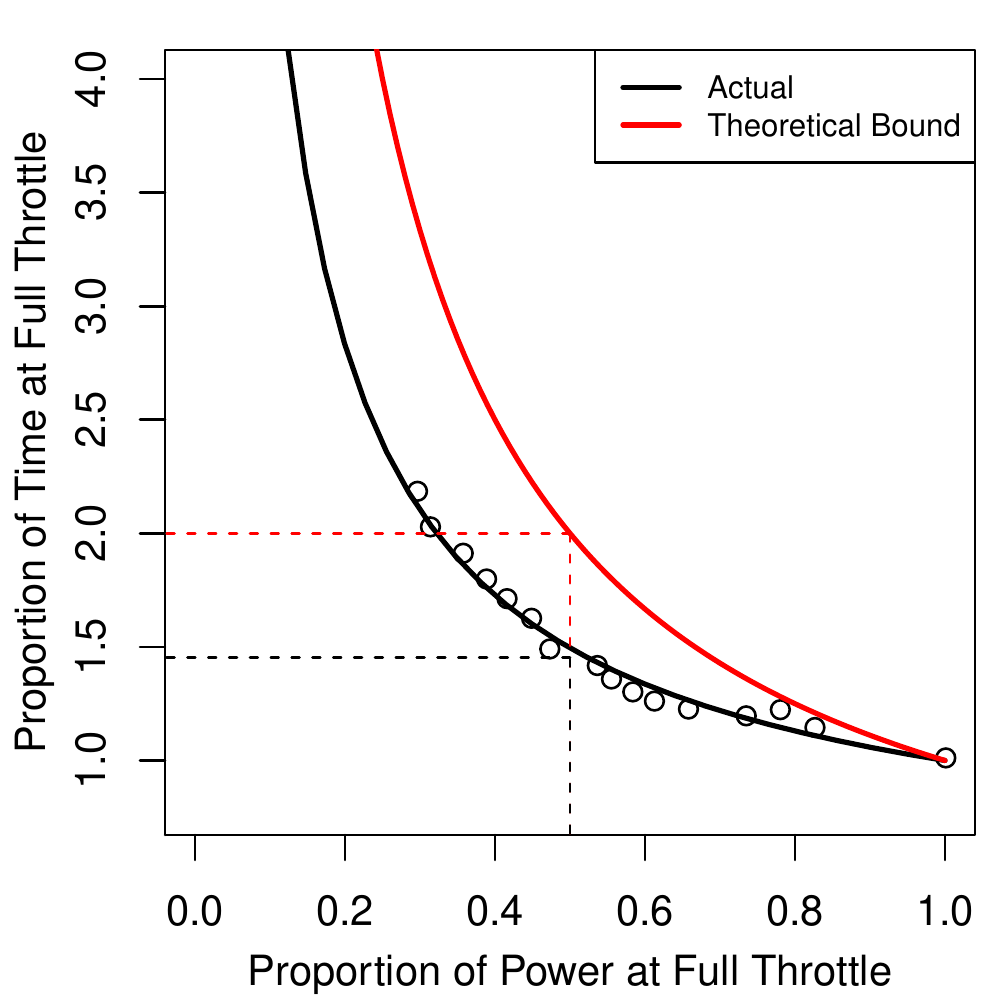}}}
\end{minipage}
\end{center}
\vspace{-.35in}
\end{figure}

The effect of power reduction via CPU throttling and its effect on performance has been previously investigated \citep{Hsu2005:power-rts,Freeh2007:energy-time,Ge2007:power-speedup,Pakin13b}.  It is known and was demonstrated in \cite{Pakin13b}, on several benchmark programs, that power scales like CPU frequency squared.  
%Figure~\ref{throttle_data} shows the quadratic effect that reducing CPU frequency has on power for a particular job.  At CPU frequency 0, the intercept, we obtain the idle power draw for a node (or cage in this case).  If a given job is 100\% compute bound, then reducing the cpu frequency by 10\% would cause that job to take 10\% longer time.  However, this is a worst case senario, as a 10\% cpu freuqnency reduction for a job that needs to use a lot of memory will not take 10\% longer.  {\bf (this paragraph above could use help)}  
It is beyond the scope of this paper to develop a precise relationship between performance and power capping.  Instead the simple logic portrayed in Figure~\ref{fig:derivation_of_bound} can be used to develop an upper bound on the performance degradation (i.e., the amount of additional time needed to complete the job) due to power capping.  

Figure~\ref{fig:derivation_of_bound}(a) displays Power by CPU frequency for the \emph{POP} program, one of the benchmarks used in \cite{Pakin13b}. The POP program, along with the other programs used in that study, runs at a relatively constant power since it is essentially performing a single task
%, i.e., POP solves a pre-specified elliptic equation using a preconditioned conjugate gradient (PCG) solver \citep{Dukowicz94}.
It is unlike a production job in that sense, since production jobs will cycle between different tasks and write output to disk at checkpoints, etc.  However, POP would mimic the behavior exhibited by a production job during one of these homogeneous tasks.  POP is only displayed here to help illustrate the general bound on performance degradation, i.e., Figure~\ref{fig:derivation_of_bound} provides a means to bound performance degradation for any program, not just POP.

Figure~\ref{fig:derivation_of_bound}(a) provides an upper bound on the power draw as a function of CPU frequency.  This bound is based on the fact that power for a node (above and beyond idle power) scales like CPU frequency squared.  It also assumes that the idle power draw of a node is constant, which is reasonable as the idle power draw of a node (or cage in this case) is easy to measure and is relatively constant.
Thus, an upper bound on power is a straight line from idle draw $I$ at CPU frequency $\chi=0$, to the power needed, $P_{\mbox{\scriptsize max}}$, when CPU frequency is at full throttle, i.e., $\chi=3.0$ GHz in this case.  Figure~\ref{fig:derivation_of_bound}(b) shows the inverse of the relationship of that in Figure~\ref{fig:derivation_of_bound}(a) on a proportional scale. That is, the proportion of full throttle CPU is given as a function of proportion of full throttle CPU power (i.e., above and beyond idle power), for both the POP program and the bound (which is now a lower bound for CPU frequency as a function of power).  Thus, if power is to be reduced by 50\%, then CPU frequency could still be allowed to be (at least) 50\% of full throttle.
%For the POP program in particular, CPU frequency could be $\sim$60\% to obtain a 50\% power reduction.
Figure~\ref{fig:derivation_of_bound}(c) shows the run time by CPU frequency for the POP program.  Figure~\ref{fig:derivation_of_bound}(d) provides the proportional multiplier of the run time at full throttle CPU as a function of CPU frequency (relative to full throttle CPU).  For a completely compute bound program (i.e., simply executing instructions while all memory is in local cache), the proportional time to completion would scale like the inverse of the CPU frequency (e.g., if the CPU frequency is reduced by 50\% then the program would take 2$\times$ as long to complete).  This is the upper bound illustrated by the red curve; most production jobs will not be 100\% compute bound at any given time. POP is close to being 100\% compute bound, but if CPU frequency is reduced by 50\% it takes POP only $\sim1.8\times$ (i.e., not 2$\times$) as long to complete.

Finally, Figure~\ref{fig:derivation_of_bound}(e) combines the relationships depicted in Figures~\ref{fig:derivation_of_bound}(b)~and~\ref{fig:derivation_of_bound}(d) to produce the relation for the proportional multiplier of run time as a function of the proportion of the power (needed at full CPU throttle) that is available.  The resulting upper bound is simply an inverse relationship, i.e., reduce power by 50\% of what the uncapped program would draw and the program would take at most 2$\times$ as long to run.  Specifically, for the POP program, if the power were reduced by 50\% of the full throttle power, then only $\sim1.5\times$ the unrestricted time would actually be needed to complete the task.

%There are other possible means to limit power to a node beyond CPU throttling.
%%, and the exact manner in which this will be done (by Trinity or future machines) has not yet been determined.
%However, the degradation to job performance would necessarily come via a CPU performance reduction.
%%\cbs{(Scott/Mike: Is this true??  If not what do I want to say here?)}.
%Thus, the bound on performance degradation provided here, i.e., assuming all power savings come at the cost of a CPU throttle down, would remain valid.
%%Certainly, there would be some benefit to further exploring the performance by power relationship for the population of jobs running on a given cluster.  However, that is beyond the scope of this paper, and the bound on performance degradation used here is not all that conservative as seen in Figure~\ref{fig:derivation_of_bound}.

\begin{wrapfigure}{r}{.5\textwidth}
\vspace{-.32in}
\begin{center}
\caption{Illustration of the Upper Bound on Performance Degradation.}
\vspace{-.2in}
\includegraphics[width=0.5\textwidth,height=0.2\textheight]{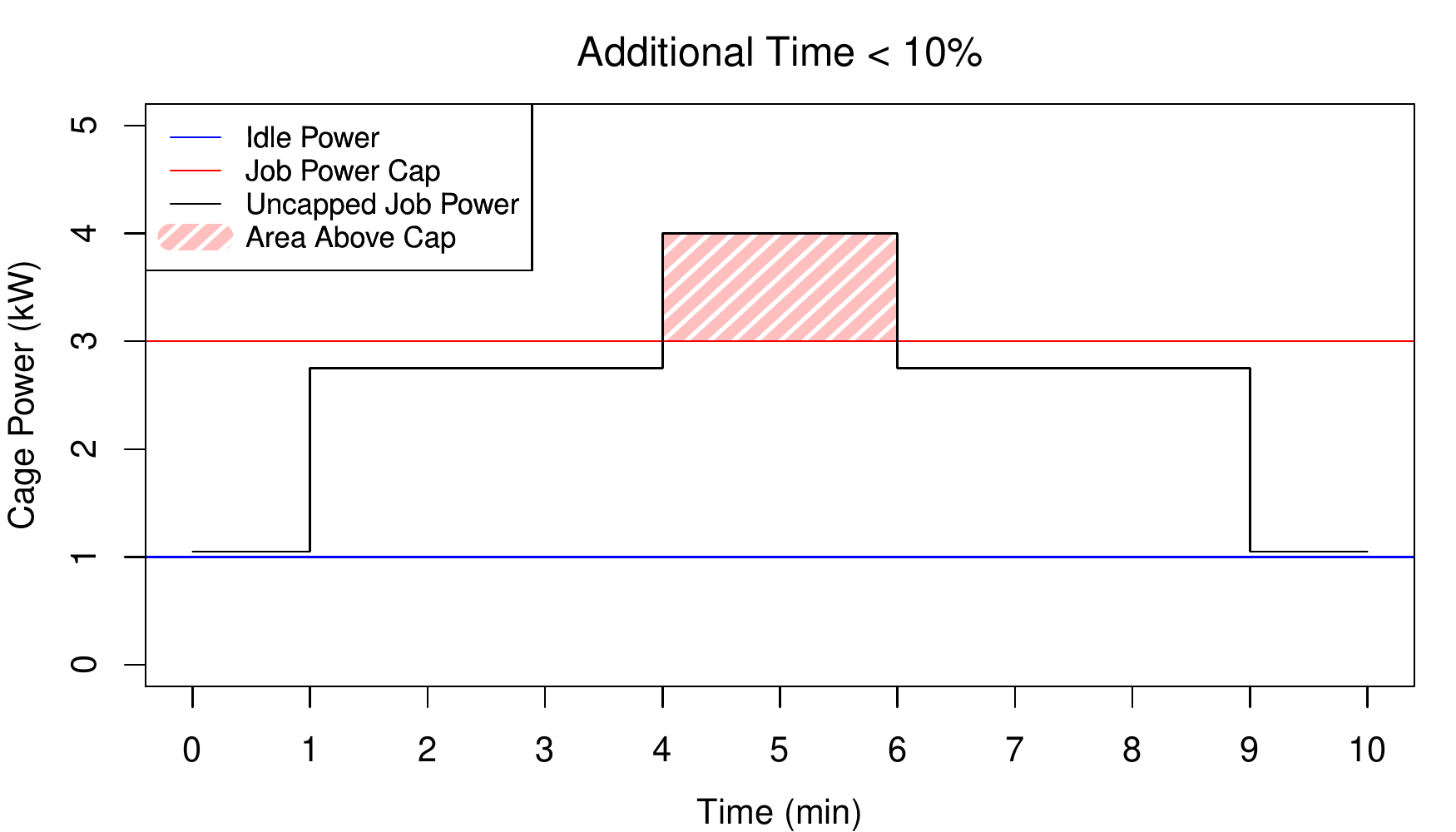}
\label{fig:job_pow_cap}
\end{center}
\vspace{-.55in}
\end{wrapfigure}

The upper bound relationship displayed in Figure~\ref{fig:derivation_of_bound}(e) can be extended to provide a bound on performance degradation for a more heterogeneous program that does not run at constant power.  Consider the program depicted in Figure~\ref{fig:job_pow_cap}, where it is known what the unrestricted/uncapped power draw would be over time.  If a power cap were hypothetically introduced at 3~kW, in this case, the two minutes of computation required to complete the task between 4 and 6 minutes in Figure~\ref{fig:job_pow_cap} would be increased.  The program wants 4~kW during that portion of the computation, but it is restricted to 3~kW, while idle draw is 1~kW, i.e., it is allowed only 2/3 of the power (above idle) that it needs during that time.  Thus, according to the bound displayed in Figure~\ref{fig:derivation_of_bound}(e) and discussed above, the program would require at most $1/(2/3)=1.5\times$ as long to complete the task.  In other words, that two minute task (at full power) would now require at most 3 minutes.  The rest of the 8 minutes of the program's execution remains unaffected by the 3~kW cap.  Thus, the program would take at most $8+3=11$ minutes to execute instead of 10 minutes, a 10\% increase in run time for the entire program.

\begin{figure}[t!]
\vspace{-.1in}
\begin{center}
\caption{Cage power over time for three distinct jobs with a power cap leading to an upper bound of 0.5\% additional time to completion.}
\vspace{-.15in}
\includegraphics[width=.85\textwidth]{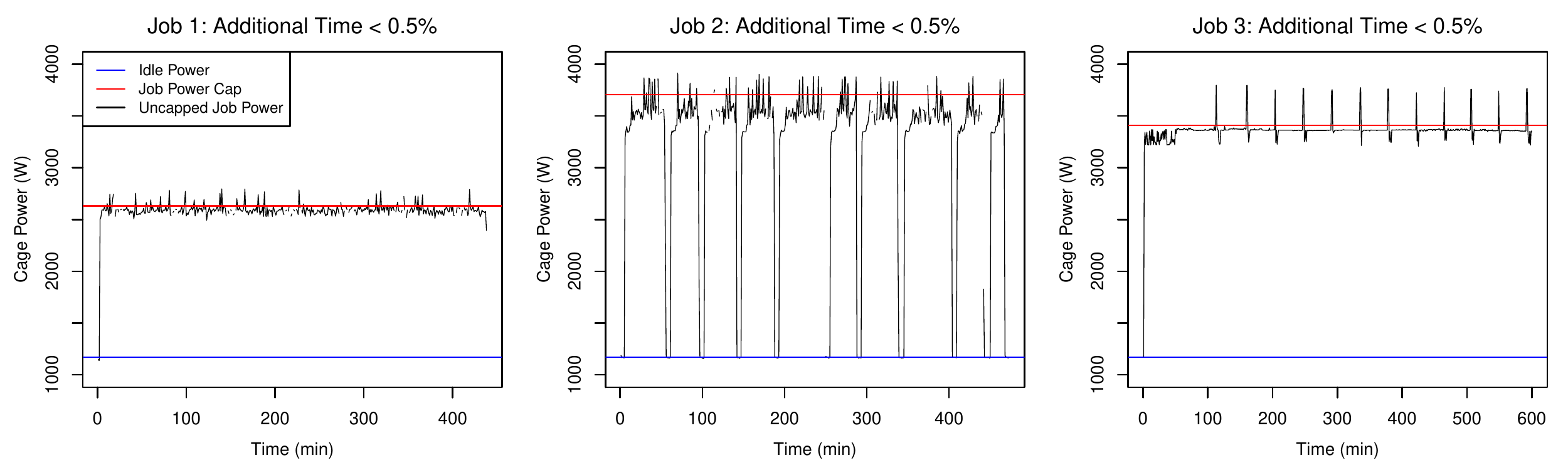}
\label{fig:3_job_pow_cap}
\end{center}
\vspace{-.35in}
\end{figure}

In general, suppose the unrestricted power profile over time $t$, for a given program, is known to be $P(t)$ and idle power is $I$.  Using the same logic as in the preceding paragraph with a quadrature argument, the increase in time $\Delta$ resulting from a power cap of $C$ would be
\vspace{-.1in}\beq
\Delta \leq \frac{\int \left| P(t) - C \right|_+ dt}{C-I},
\vspace{-.1in}\eeq
where $|x|_+ = x$ if $x>0$ and zero otherwise.
This relationship is used in Figure~\ref{fig:3_job_pow_cap} to find the power cap that would produce at most a 0.5\% increase in computation time for the three job examples from Figure~\ref{fig:3_job_pow}.  The cap leading to this 0.5\% increase is much lower for Job 1 than for the other two jobs.  The fact that this cap could be a lot lower for some jobs than for others is really the driving force of this paper.  The goal is to set the same cap for a group of nodes runnning the same job, in a manner such that no job receives a major performance degradation.  In reality, however, the unrestricted power that a job will draw will not be known ahead of time.
%, making it impossible to provide an optimal capping strategy for the current job mix on a machine directly as above.
Thus, we next propose a statistical model for job power draw and use this model to provide an approximate predictive distribution of the future power profile for each job running on the machine.  From this probabilistic description of the $P(t)$ for each job, one can devise an informed capping strategy to optimize a given criterion.

\vspace{-.2in}
\section{Hierarchical Bayesian Model for Job Power}
\vspace{-.1in}
\label{sec:Bayes_job_power}

%As discussed previously, a hierarchical model is assumed for the node-level power draw for a job.
The model for an individual job is described in Section~\ref{sec:job_model}, and then the parent model governing the parameters of the individual job model is introduced in Section~\ref{sec:parent_model}.  Estimation of the parent model parameters and then the individual job parameters (i.e., updating) is then discussed, both from a fully Bayesian and a simple, pragmatic perspective.

\vspace{-.15in}
\subsection{Statistical Model for an Individual Job}
\vspace{-.05in}
\label{sec:job_model}

The model for the power drawn for a given job is a hidden Markov model that allows a job to switch between various regimes (or tasks, e.g., see Figure~\ref{fig:3_job_pow}) and draw a different amount of power on average while performing each task.  Specifically, the power draw observation for the $j\th$ job, $j=1,\dots,J$ at time $t$ is represented as
\vspace{-.14in}\beq
x_j(t)=\sum_{k=1}^\infty \mu_{j,k} I_{\{\xi_j(t)=k\}} + z_j(t) + \eps_j(t),
\label{eq:job_pow}
\vspace{-.14in}\eeq
where (i) $\xi_j(t)$ is an indicator for a particular regime (i.e., task that is being performed) during the job, (ii) $\mu_{j,k}$ is the mean level power draw while job $j$ is in the $k\th$ regime, (iii) $z_j(t)$ is a stationary, mean zero, time dependent process that allows the power to fluctuate around the current mean level, and (iv) $\eps_j(t) \stackrel{iid}{\sim} N(0, \ttau^2)$ is a white noise measurement error, with common variance for all jobs.  For simplicity in the estimation procedure, we assume that $t$ is discrete, since the power observations are recorded at regular one-minute intervals, $t=1,2,\dots,T_j$.  However, a continuous time analog is implied by the following discrete time model if, for example, predictions on a finer grid than every minute were desired.

From exploratory analysis of a homogeneous task (i.e., the POP program and similar), it was deemed appropriate to model $z_j(t)$ as an Ornstein-Uhlenbeck (O-U) process, i.e., 
\vspace{-.13in}\beq
z_j \sim GP\left(0, \sigma_j^2\Gamma_{\rho_j} \right),
\label{eq:zj_model}
\vspace{-.13in}\eeq
where GP denotes a Gaussian Process and $\Gamma_{\rho}(s,t)=\exp\{-\rho |s-t|\}$.  This is equivalent to a first order auto-regressive model in regularly-spaced discrete time.

Finally, the regime indicator process $\xi_j(t)$ is assumed to be a Markov chain where the residence time in regime $\xi_j(t)=k$ is
\vspace{-.13in}\beq
T \sim \mbox{Geometric}(\lambda_{j,k}(1-\pi_{j,k})).
\label{eq:trans_time}
\vspace{-.13in}\eeq
The parametrization of the geometric rate with $\lambda_{j,k}(1-\pi_{j,k})$ above may seem redundant and unidentifiable at first glance.  However, the characterization of state transition probabilities below identifies these parameters.  The reason for the parametrization in (\ref{eq:trans_time}) is that it allows for conjugate updates of the $\pi_{j,k}$ (see the Supplementary Material), while allowing for essentially the same model as if only $\lambda_{j,k}$ were used for the geometric rate.

Suppose that a transition out of state $k$ occurs at time $u$, then it is assumed that
\vspace{-.1in}\beq
\Pr(\xi_j(u)=l \mid \xi_j(u-1)=k) \propto \left\{
\begin{array}{ll}
  \pi_{j,l}  & \mbox{for} \; l\neq k \\
  0 & \mbox{otherwise}.
\end{array}
\right.
\label{eq:xi_transition_probs}
\vspace{-.1in}\eeq
That is, when a transition occurs, a \emph{new} regime is chosen with probability proportional to  $\bpi_j=[\pi_{j,1},\pi_{j,2},\dots]'$, regardless of the state $k$ from which the transition is being made.

The model for $\xi_j$ above is a discrete time Markov chain (MC), but was intentionally parameterized analogously to a continuous time MC, (i.e., geometric in place of exponential residence time, then it moves on to a new regime).  In (\ref{eq:xi_transition_probs}), although the transition rates are assumed independent of the previous state, the residence times in each state are allowed different $\lambda_{j,k}$.  Thus, this parametrization results in a transition probability matrix (TPM) for the discrete time MC with unequal rows in general.  In particular, the diagonal of the TPM will generally be inflated, i.e., the $(k,l)\th$ element of the TPM is,
\vspace{-.1in}\beq
P_{k,l} = \Pr\left(\xi_j(t+1) = l \mid \xi_j(t) = k \right) = \left\{
\begin{array}{ll}
  \lambda_{j,k} \pi_{j,l} & \mbox{for $l \neq k$,} \\
  \lambda_{j,k} \pi_{j,k} + \left(1- \lambda_{j,k} \right) & \mbox{for $l = k$.}
\end{array}
\right.
\label{eq:xi_TPM}
\vspace{-.1in}\eeq
The transition model in (\ref{eq:xi_TPM}) is related to the sticky HDP-HMM of \cite{Fox:2011} who allow for a distinct set of transition probabilities from each current state.  They consider a single chain, so we drop the dependence on $j$, i.e., $\mbox{Prob}(\xi(t)=l | \xi(t-1) =k) = \pi_{kl}$.  The transition probabilities for each state, $\pi_k = (\pi_{k1}, \pi_{k2}, ...)$, are each modeled with a DP with inflated prior mass on the diagonals $\pi_{kk}$ to encourage the process to stay in the current state.  They use a HDP to pool information across states to estimate the transition probabilities.

The rationale for the proposed model in (\ref{eq:xi_transition_probs}) and (\ref{eq:xi_TPM}) is that the distribution of the residence time in a given regime can vary greatly between the regimes within the same job (refer back to Figure~\ref{fig:3_job_pow}, for example).  However, upon \emph{leaving} a regime, it was not immediately clear from the data that the next regime was dependent on the previous regime.  Most programs have only a few transitions from each of their \emph{observed} regimes, which would make estimation of completely separate transition probabilities difficult for practical purposes.
%For a real job, the regime switching process is not even a MC at all, as it is largely deterministic (i.e., which task is computed next, when to write out a check point to disk, etc.).  However, the logic and hardware involved are complicated enough to make it look somewhat random.
In any case, the proposed model adequately represents the regime changes for our purposes, see Section~\ref{sec:job_CV}.

There are 454 observations in the data set, but only 213 unique jobs.  That, is, some jobs span multiple cages.  In these cases, the jobs are assumed to share common parameters $\mu_{j,k}$, $\sigma_j^2$, $\rho_j$, $\lambda_{j,k}$, and $\pi_{j,k}$, but the values of the random processes $\xi_j$, $z_j$, and $\eps_j$ are treated as independent across replicates.  Initial inspection of several replicate job observations, implied that the regimes were not quite in lock step with one another over time so that allowing for separate $\xi_j$ was necessary.  Still, it could be beneficial for estimation purposes to model the offset of the $\xi_j$ for replicate jobs, i.e., introduce a dependence between their respective $\xi_j$.  For the purpose of prediction of the entire machine in Section~\ref{sec:analysis}, it will be assumed that jobs spanning multiple cages (or ultimately nodes) do have identical $\xi_j$ processes, which would produce a conservative prediction of the aggregate power drawn by such a job, e.g., all nodes running the same job would be assumed to be in the most power intensive regime at the same time.

\vspace{-.15in}
\subsection{Parent Model for the Job Parameters}
\vspace{-.05in}
\label{sec:parent_model}

There are several parameters in the model for a given job in Section~\ref{sec:job_model}, e.g., $\mu_{j,k}$, $\lambda_{j,k}$, $\sigma_{j}$, etc.  Further, we wish to be able to make predictions, even for newly started jobs with very little or no data.  A typical approach in such cases is to assume that job specific parameters come from a parent distribution which is described below.
%In the Bayesian paradigm, the parent parameters governing the parent distribution have prior distributions placed on them, and their posterior distribution is approximated via MCMC.
%Here we describe the details of the parent model, and defer the computational details to Section~\ref{sec:DPM_estimation}.

\vspace{.1in}
\noindent
{\bf \em Parent Model for $z_j$ Parameters}.  In the model for the power fluctuations $z_j$ in (\ref{eq:zj_model}), there are two job specific parameters, $\sigma_j^2$ and $\rho_j$.  We assume these parameters for the $j\th$ job are realized from a parent distribution as follows,
\vspace{-.15in}\begin{eqnarray}
\sigma_j^2 & \stackrel{iid}{\sim}  & \mbox{log}N(\tmu_\sigma, \tsigma^2_\sigma), \; j=1,\dots,J, \nonumber \\
\rho_j & \stackrel{iid}{\sim}  & \mbox{log}N(\tmu_\rho, \tsigma^2_\rho), \; j=1,\dots,J,  \nonumber \\[-.45in] \nonumber
\end{eqnarray}
where $\mbox{log}N$ is the log-Normal distribution.  Here and throughout the paper, any parameters that are parent parameters %(i.e., shared across jobs)
receive a tilde above them in their notation to add clarity.  A log-Normal distribution was chosen for $\sigma_j^2$ as opposed the popular conjugate choice of Inverse-Gamma (IG), due to the fact that the IG would have far too heavy of a tail to adequately represent the parent distribution of $\sigma_j^2$ variation among jobs.  Power predictions of a \emph{brand-new} job, for example, would be allowed to be significantly higher than realistic limits if using an IG model for $\sigma_j^2$.

To make the model specification complete, a prior distribution is placed on the parent parameters $\tmu_\sigma$, $\tsigma^2_\sigma$, $\tmu_\rho$, and $\tsigma^2_\rho$.  These parameters are assumed to be distributed as,
\vspace{-.12in}\beq
  \tmu_\sigma \sim N(M_\sigma, S^2_\sigma),\;\; 
  \tsigma^2_\sigma \sim \mbox{IG}(A_\sigma, B_\sigma),\;\; 
  \tmu_\rho \sim N(M_\rho, S^2_\rho) ,\;\; 
  \tsigma^2_\rho \sim \mbox{IG}(A_\rho, B_\rho).
  \label{eq:hyper_beta_rho}
\vspace{-.12in}\eeq
%\vspace{-.5in}\begin{eqnarray}
%  \tmu_\sigma & \sim & N(M_\sigma, S^2_\sigma) \nonumber \\
%  \tsigma^2_\sigma & \sim & \mbox{IG}(A_\sigma, B_\sigma) \nonumber \\
%  \tmu_\rho & \sim & N(M_\rho, S^2_\rho) \nonumber \\
%  \tsigma^2_\rho & \sim & \mbox{IG}(A_\rho, B_\rho), \label{eq:hyper_beta_rho}
%  \\[-.55in] \nonumber
%\end{eqnarray}
Hyper-prior parameters (e.g., $M_\sigma, S^2_\sigma$) for all parent parameter prior distributions defined in (\ref{eq:hyper_beta_rho}) and below are always denoted by capital letters and a corresponding subscript.  Values must be set for all such parameters in order to complete the model specification.  For convenience, all of these parameters will be reviewed and specified for this application at the end of this section.

\vspace{.1in}
\noindent
{\bf \em Parent Model for $\mu_{j,k}$}. The mean level $\mu_{j,k}$ for the $k\th$ regime of job $j$ in (\ref{eq:job_pow}) is assumed to come from a (possibly infinite) normal mixture model, i.e., 
\vspace{-.1in}\beq
\mu_{j,k} \stackrel{iid}{\sim} \sum_{m=1}^\infty \tomega_{m} N(\tnu_{m}, \tvarsigma_{m}^2), \; \mbox{$k=1,2,\dots$, and $j=1,\dots,J$.}
\label{eq:mujk_model}
\vspace{-.1in}\eeq
where $\sum \tomega_m = 1$.
The normal mixture model in (\ref{eq:mujk_model}) is assumed to be a Dirichlet process \citep{Ferguson73,Ishwaran01,LidHjort10}.  That is, the mixture probabilities follow a stick-breaking distribution \citep{Sethuraman94}, $\btomega=[\tomega_{1},\tomega_{2},\dots]' \sim \mbox{SB}(\tgamma)$, or
\vspace{-.175in}\beq
\tomega_{m} = u_m \prod_{n=1}^{m-1} (1-u_n)
\label{eq:u_m}
\vspace{-.15in}\eeq
where $u_m \stackrel{iid}{\sim} \mbox{Beta}(1,\tgamma)$, $m=1,2,\dots$.  A further hyper-prior is typically assumed on $\tgamma$, i.e.,
\vspace{-.12in}\beq
\tgamma  \sim \mbox{Gamma}(A_\gamma, B_\gamma).
\label{eq:hyper_gamma}
\vspace{-.15in}\eeq

The remaining parent parameters for $\mu_{j,k}$ distribution are assumed to have the following prior distributions,
\vspace{-.3in}\begin{eqnarray}
\tnu_{m} & \stackrel{iid}{\sim} & N(M_\nu, S^2_\nu), \; m=1,2,\dots, \nonumber \\
\tvarsigma_{m}^2 & \stackrel{iid}{\sim} & \mbox{IG}(A_\varsigma, B_\varsigma), \; m=1,2,\dots \label{eq:hyper_varsigma}
  \\[-.5in] \nonumber
\end{eqnarray}

\vspace{.1in}
\noindent
{\bf \em Parent Model for $\xi_j$ Parameters}.  The $\xi_j$ process is governed by the parameters $\lambda_{j,k}$ and $\pi_{j,k}$, i.e., the regime transition rate and transition probabilities from (\ref{eq:trans_time}) and (\ref{eq:xi_transition_probs}), respectively.  The $\lambda_{j,k}$ in the parametrization of the regime transition rates for each job are assumed to be,
\vspace{-.12in}\beq
\lambda_{j,k} \stackrel{iid}{\sim} \mbox{Beta}(\talpha_\lambda, \tbeta_\lambda), \; \mbox{$k=1,2,\dots$, and $j=1,\dots,J$.} \nonumber
%\label{eq:hyper_lambda}
\vspace{-.12in}\eeq
Similar to the model for $\tomega_{m}$ in (\ref{eq:u_m}), it is assumed that the transition probabilities for $\xi_j$ in (\ref{eq:xi_transition_probs}) come from a stick-breaking distribution.  That is, $\bpi_j=[\pi_{j,1},\pi_{j,2},\dots]' \sim \mbox{SB}(\tdelta)$, or
\vspace{-.12in}\beq
\pi_{j,k} = v_{j,k} \prod_{l=1}^{k-1} (1-v_{j,l})
\label{eq:v_jk}
\vspace{-.12in}\eeq
where $v_{j,k} \stackrel{iid}{\sim} \mbox{Beta}(1,\tdelta)$, $k=1,2,\dots$, and $j=1,\dots,J$.

The parent parameters $\talpha_\lambda$, $\tbeta_\lambda$, and $\tdelta$ have the following prior distribution,
\vspace{-.12in}\beq
\talpha_\lambda  \sim  \mbox{Gamma}(A_\lambda, B_\lambda),\;\;  
\tbeta_\lambda  \sim  \mbox{Gamma}(C_\lambda, D_\lambda), \;\;
\tdelta  \sim  \mbox{Gamma}(A_\delta, B_\delta).
\label{eq:hyper_delta}
\vspace{-.13in}\eeq
%\vspace{-.3in}\begin{eqnarray}
%\talpha_\lambda & \sim & \mbox{Gamma}(A_\lambda, B_\lambda),  \nonumber \\
%\tbeta_\lambda & \sim & \mbox{Gamma}(C_\lambda, D_\lambda), \nonumber \\
%\tdelta & \sim & \mbox{Gamma}(A_\delta, B_\delta).
%\label{eq:hyper_delta}
%  \\[-.55in] \nonumber
%\end{eqnarray}

\vspace{.1in}
\noindent
{\bf \em Prior Distribution for $\ttau^2$}. Lastly, the measurement error variance $\ttau^2$ of the model in (\ref{eq:job_pow}) is common to each job and is assumed to have prior distribution,
\vspace{-.13in}\beq
\ttau^2 \sim \mbox{IG}(A_\tau, B_\tau)
\label{eq:hyper_tau}
\vspace{-.17in}\eeq

\vspace{.1in}
\noindent
{\bf \em Summary of the Hierarchical Model Parameters and Prior Specification}.
Table~\ref{tab:priors} summarizes the hierarchical model and provides the values used in the prior specifications.  Relatively diffuse priors were used for most parameters except the regime location distribution, the measurement error variance, and the O-U process parameters.  Priors for these values were formulated based on data from the performance study of \cite{Pakin13b}.

\begin{table}[h!]
  \vspace{-.11in}
\setlength{\tabcolsep}{4pt}
\begin{center}
  \caption{Summary of hierarchical model and the specification of the parent prior distributions.}
\label{tab:priors}
\vspace{-.07in}
{\small
\begin{tabular}{|l|c|c|c|}
  \hline
  Description & Job Parameter Model & Parent Prior & Specification \\
  \hline
  \multirow{4}{*}{Variance of the O-U Process $z_j$} &  \multirow{4}{*}{$\sigma_j^2 \stackrel{iid}{\sim} \mbox{log}N(\tmu_\sigma, \tsigma^2_\sigma)$} & \multirow{2}{*}{$\tmu_\sigma \sim N(M_\sigma, S^2_\sigma)$} & $M_\sigma= 4$ \\
  & & & $S^2_\sigma=1$ \\
  \cline{3-4}
  & & \multirow{2}{*}{$\tsigma^2_\sigma \sim \mbox{IG}(A_\sigma, B_\sigma)$} & $A_\sigma=10$ \\
  & & & $B_\sigma=5$ \\
  \hline
%%%%%%%%%%%%%%%%%%%%%%%%%%%%%%%%%%%%%%%%%%%%%%%%%%%%%%%%%%%%%%%%%%%%%%%%%%%%
  \multirow{4}{*}{Range of the O-U Process $z_j$} & \multirow{4}{*}{$\rho_j \stackrel{iid}{\sim} \mbox{log}N(\tmu_\rho, \tsigma^2_\rho)$} & \multirow{2}{*}{$\tmu_\rho \sim N(M_\rho, S^2_\rho)$} & $M_\rho= -2$ \\
  & & & $S^2_\rho=9$ \\
  \cline{3-4}
  & & \multirow{2}{*}{$\tsigma^2_\rho \sim \mbox{IG}(A_\sigma, B_\sigma)$} & $A_\sigma=10$ \\
  & & & $B_\sigma=5$ \\
  \hline
%%%%%%%%%%%%%%%%%%%%%%%%%%%%%%%%%%%%%%%%%%%%%%%%%%%%%%%%%%%%%%%%%%%%%%%%%%%%
  \multirow{6}{*}{Location of the $k\th$ regime} & \multirow{6}{*}{$\mu_{j,k} \stackrel{iid}{\sim} \sum  \tomega_{m} N(\tnu_{m}, \tvarsigma_{m}^2)$} & $\tomega_{m} \sim \mbox{SB}(\tgamma)$, & $A_\gamma= 1$ \\
  & & $\tgamma  \sim \Gamma(A_\gamma, B_\gamma)$ & $B_\gamma=1$ \\
  \cline{3-4}
  & & \multirow{2}{*}{$\tnu_{m} \stackrel{iid}{\sim}  N(M_\nu, S^2_\nu)$} & $M_\nu=2000$ \\
  & & & $S^2_\nu=10^6$ \\
  \cline{3-4}
  & & \multirow{2}{*}{$\tvarsigma_{m}^2 \stackrel{iid}{\sim} \mbox{IG}(A_\varsigma, B_\varsigma)$} & $A_\varsigma=1$ \\
  & & & $B_\varsigma=1$ \\
  \hline
%%%%%%%%%%%%%%%%%%%%%%%%%%%%%%%%%%%%%%%%%%%%%%%%%%%%%%%%%%%%%%%%%%%%%%%%%%%%
  \multirow{4}{*}{Transition rate for $k\th$ regime} &  \multirow{4}{*}{$\lambda_{j,k} \stackrel{iid}{\sim} \mbox{Beta}(\talpha_\lambda, \tbeta_\lambda)$} & \multirow{2}{*}{$\talpha_\lambda \sim \Gamma(A_\lambda, B_\lambda)$} & $A_\lambda= 1$ \\
  & & & $B_\lambda=1$ \\
  \cline{3-4}
  & & \multirow{2}{*}{$\tbeta_\lambda \sim \Gamma(C_\lambda, D_\lambda)$} & $C_\lambda= 1$ \\
  & & & $D_\lambda=1$ \\
  \hline
%%%%%%%%%%%%%%%%%%%%%%%%%%%%%%%%%%%%%%%%%%%%%%%%%%%%%%%%%%%%%%%%%%%%%%%%%%%%
  \multirow{2}{*}{Regime transition probabilities} &  \multirow{2}{*}{$\bpi_{j} \stackrel{iid}{\sim} \mbox{SB}(\tdelta)$} & \multirow{2}{*}{$\tdelta \sim \Gamma(A_\delta, B_\delta)$} & $A_\delta= 1$ \\
  & & & $B_\delta=1$ \\
  \hline
%%%%%%%%%%%%%%%%%%%%%%%%%%%%%%%%%%%%%%%%%%%%%%%%%%%%%%%%%%%%%%%%%%%%%%%%%%%%
  \multirow{2}{*}{Observation Error Variance} & \multirow{2}{*}{$-$}  & \multirow{2}{*}{$\ttau^2 \sim \mbox{IG}(A_\tau, B_\tau)$} & $A_\tau= 10$ \\
  & & & $B_\tau=10$ \\
  \hline
\end{tabular}
}
\end{center}
\vspace{-.28in}
\end{table}

\vspace{-.15in}
\subsection{Estimation of Model Parameters}
\vspace{-.05in}
\label{sec:DPM_estimation}

Complete MCMC details, including full conditional distributions, etc., are provided in the Supplementary Material.  However, an overview is provided here to illustrate the main idea.  The MCMC routine is a typical hybrid Gibbs, Metropolis Hastings (MH) sampling scheme (e.g., see \cite{Givens05}).
Each MCMC iteration consists of the following two steps:
\vspace{-.06in}
\begin{itemize}
\item[(i)] Update job-specific parameters for each job, conditional on the parent parameters.\\[-.375in]
\item[(ii)] Update parent parameters, conditional on the job-specific parameters from (i).
\end{itemize}
\vspace{-.05in}
Conditional on the parent parameters, the parameters for each job are independent across job, making the many (213 in this case) job specific updates easily parallelizable.

The job specific parameters sampled in the MCMC are 
\vspace{-.14in}\bdm
\Theta_j = \left\{
\left\{\xi_j(t)\right\}_{t=1}^{T_j},
\left\{\lambda_{j,k}\right\}_{k=1}^K,
\left\{\pi_{j,k}\right\}_{k=1}^K,
\left\{\mu_{j,k}\right\}_{k=1}^K,
\left\{z_j(t)\right\}_{t=1}^{T_j},
\sigma_j^2,
\rho_j
\right\}, \; j=1,\dots,J.
\vspace{-.14in}\edm
For convenience of computation, the number of components in stick-breaking model for $\pi_{j,k}$ was capped at a finite value $K$, i.e., $k=1,\dots,K$.  The value of $\pi_{j,K}$ was observed and $K$ was increased until $\pi_{j,K}$ values were negligible for all jobs at $K=10$.
Because of the discrete representation of the job power process, 
the job specific parameters have relatively simple conjugate updates (details provided in the Supplementary Material).  Two exceptions are $\sigma_j^2$ and $\rho_j$, which require MH updates.  However, the proposal for $\sigma_j^2$ is provided by matching the moments of the log-normal prior to an inverse-Gamma, and producing the corresponding conjugate update.  This proposal is then accepted or rejected in the usual MH fashion.  This approach resulted in $>80$\% acceptance for all $j$ along with the benefit that it requires no tuning.  The $\rho_j$ were updated via a random walk proposals.  However, the random walk was conducted on the log scale, i.e., $\log(\rho_j^*) = \log(\rho_j+\epsilon)$ for a deviate $\epsilon \sim N(0,s^2)$.  With the use of the log scale, a constant tuning parameter $s^2=0.25$ could be used for all jobs to achieve acceptances in the range of (30\% - 55\%).

The parent parameters sampled in the MCMC are
\vspace{-.14in}\bdm
\Theta^P = \left\{\tmu_\sigma, \tsigma_\sigma, \tmu_\rho, \tsigma_\rho, \left\{\tomega_m\right\}_{m=1}^M, \left\{\tnu_m\right\}_{m=1}^M, \left\{\tvarsigma^2_m\right\}_{m=1}^M, \talpha_\lambda, \tbeta_\lambda, \tdelta, \tgamma, \ttau^2 \right\}.
\vspace{-.14in}\edm
Most parent parameters have conjugate updates, with only $\alpha_\lambda$ and $\beta_\lambda$ requiring MH updates via a random walk as described for the updates of $\rho_j$ above.
The proposals were tuned to produce $\sim 40$\% acceptance in both cases.
Five different MCMC chains with varying starting points were run out to 10,000 iterations.  Based on trace plots of the parent parameters, all chains converged to the approximately the same posterior distribution
%(aside from label switching of the $m$ index for $\tomega_m$, $\tnu_m$, and $\tvarsigma^2_m$)
after about 2,000 iterations.

\vspace{-.15in}
\subsection{Updating a Given Job}
\vspace{-.05in}
\label{sec:DPM_job_update}

\begin{wrapfigure}{r}{.5\textwidth}
\vspace{-.5in}
\begin{center}
\caption{Parent normal mixture distribution for the location parameters $\mu_{j,k}$ of the regimes in an individual job process.}
\vspace{-.3in}
\includegraphics[width=0.5\textwidth]{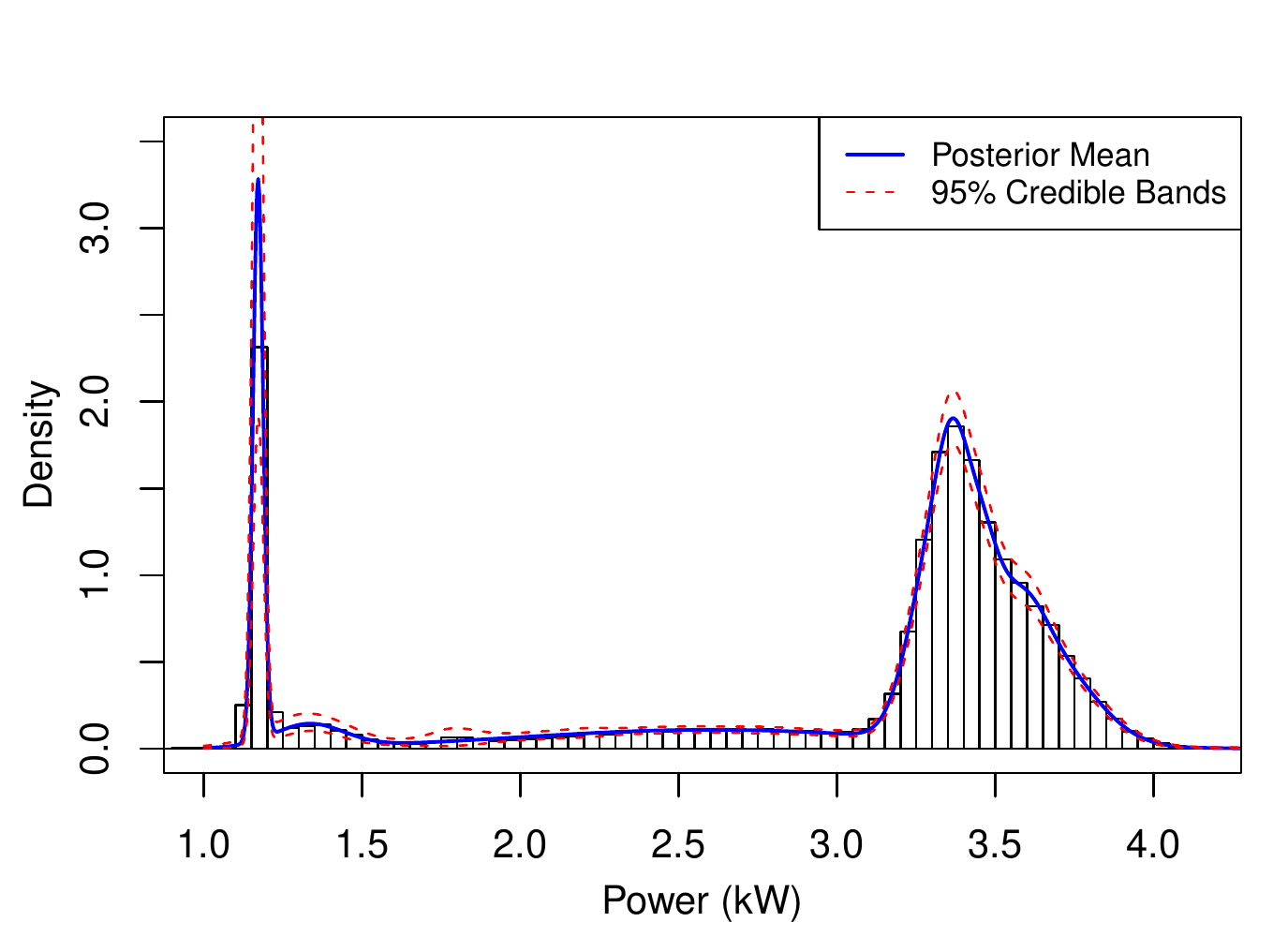}
\label{fig:parent_location}
\end{center}
\vspace{-.7in}
\end{wrapfigure}

The main goal of this work (i.e., intelligent node-level power capping) hinges on the ability to make predictions about the future power profile for a job, possibly given some previous measurements from that job.  To facilitate this goal, we leverage the fact that the uncertainty in the parent parameters is negligible relative to the uncertainty present in estimating the job parameters for a given job.
%This should be intuitively clear since all jobs in the training set contribute all of their data toward estimation of $\sim 30$ parent parameters.  And of these parameters, several of the $\tomega_m$, $\tnu_m$, $\tvarsigma^2_m$ for larger $m$ values do not matter that much as they have little influence on the overall parent density for the regime level $\mu_{j,k}$ in (\ref{eq:mujk_model}).
A posterior summary of the density for $\mu_{j,k}$ is provided in Figure~\ref{fig:parent_location}.  The solid curve is the posterior mean value of the density, while the dashed lines provide 95\% (pointwise) credible bands.  For reference a histogram is also drawn based on the sampled values of $\mu_{j,k}$ in the posterior.  The tight credible bands around the posterior mean in Figure~\ref{fig:parent_location} serve to illustrate the point above about negligible uncertainty in the parent parameters.  Thus, when updating the parameters of a specific job for prediction purposes, an empirical Bayes approach is taken where the uncertainty in the parent parameters is ignored, i.e., their values are fixed at their posterior mean.  Label-switching issues with the parameters of the normal mixture for $\mu_{j,k}$ would render their posterior mean unusable, however, the overall density for $\mu_{j,k}$ is immune to label switching.  Therefore the posterior mean density is evaluated on a fine grid as in Figure~\ref{fig:parent_location}, then approximated with a best fitting normal mixture of 10 components to provide the fixed value of the parent normal mixture parameters $\{\tomega_m$, $\tnu_m$, $\tvarsigma^2_m\}_{m=1}^{10}$.  The remaining parent parameters were fixed at their posterior mean values.

\begin{comment}

  \begin{figure}[t!]
\begin{center}
\caption{Parent normal mixture distribution for the location parameters $\mu_{j,k}$ of the regimes in an individual job process.}
\vspace{-.25in}
\includegraphics[width=.6\textwidth]{parent_level_distn_2}
\label{fig:parent_location}
\end{center}
\vspace{-.2in}
\end{figure}

\end{comment}

Once the parent parameters are fixed, a posterior distribution for the job parameters for a job given its previous power observations can be sampled by simply iterating step (i) of the MCMC algorithm in Section~\ref{sec:DPM_estimation}.  One additional caveat is that in operation, the nodes will have a power cap, making some of the observations right-censored.  However, this can easily be handled by simply sampling such observations conditional on the other parameters (and conditional on being greater than the cap) in the MCMC iterations.  In this way, the rest of the algorithm remains unchanged, as if no censoring occurred.

In practice, the many ($\sim 100$ for Luna) jobs running at a given time on a machine will need to be updated simultaneously.  However, this is once again easily parallelizable.  Because of the simplicity of the updates for each of the job parameters, the MCMC routine for a single job is very fast.  For example, to update a job that has been observed for 200 minutes requires $\sim 1$ minute for 10,000 MCMC iterations.  Convergence in most cases happens very quickly as well (within the first 1,000 iterations).  Since the intention is to make node-level power capping changes on the order of minutes, this approach is readily applicable in practice.

\begin{figure}[t]
\vspace{-.1in}
\begin{center}
\caption{Example realizations of the future after updating the 3 example jobs from Figure~\ref{fig:3_job_pow} (given 200 minutes of history). (a) Updated using uncapped data history.  (b) Updated using capped (i.e., right censored) at the $95\th$ percentile of the historical data.}
\vspace{-.1in}
\subfloat[Future realization after updating with no power cap]{\includegraphics[width=.85\textwidth]{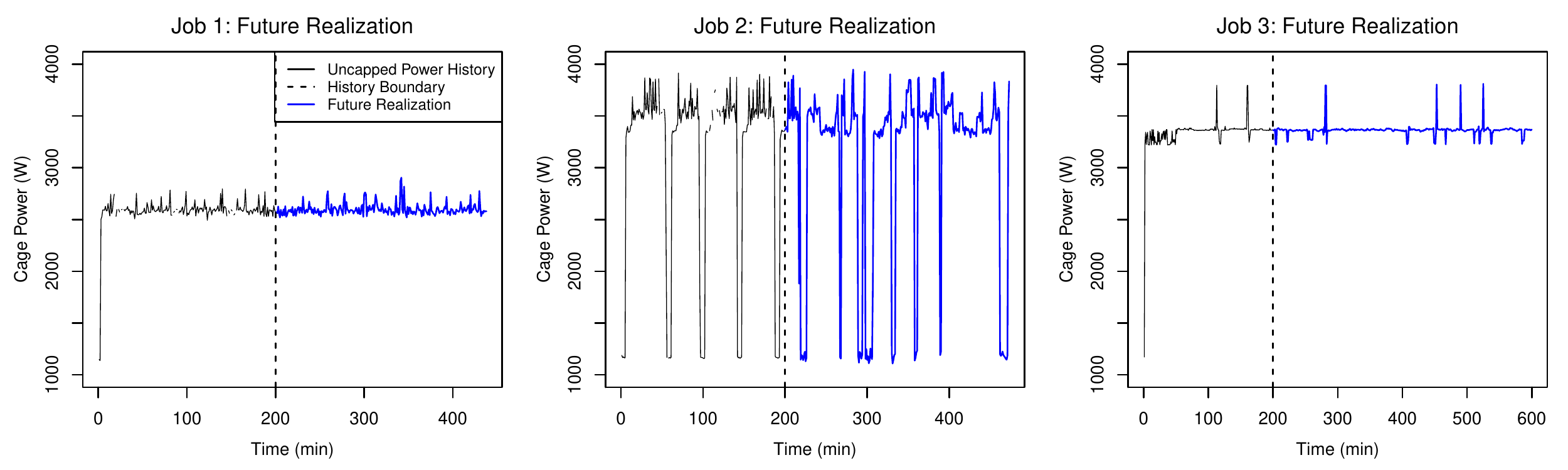}}\\[-.05in]
\subfloat[Future realization after updating with displayed power cap]{\includegraphics[width=.85\textwidth]{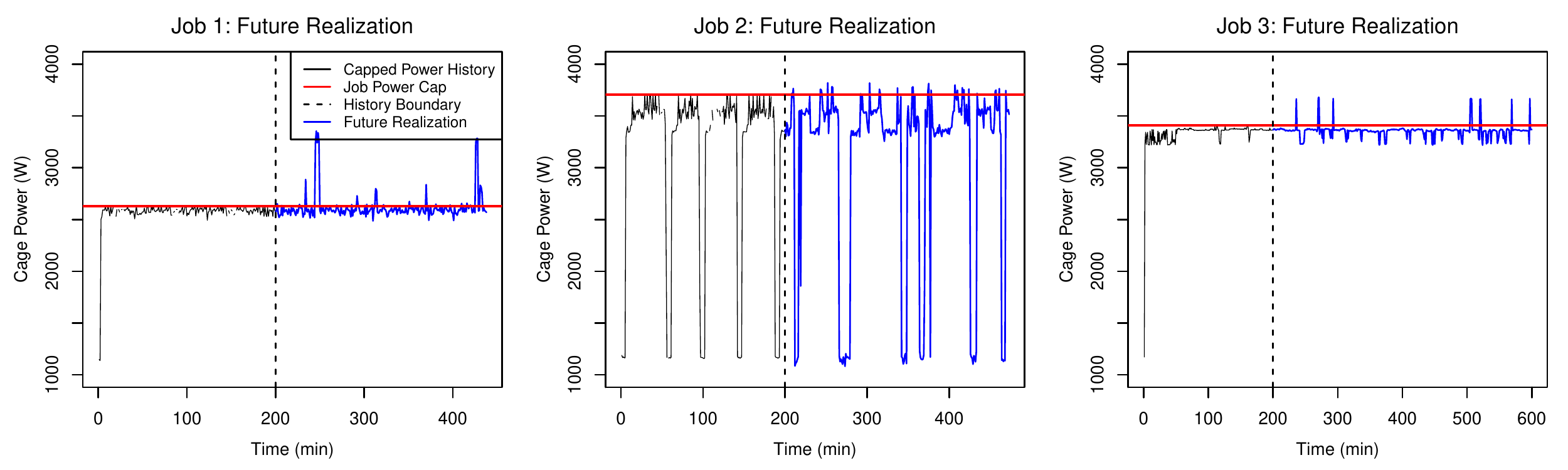}}
\label{fig:3_job_updating}
\end{center}
\vspace{-.45in}
\end{figure}

Figure~\ref{fig:3_job_updating} displays example realizations of the future power after updating the three example jobs from Figure~\ref{fig:3_job_pow} using 200 minutes of history.  The realizations for the three jobs in Figure~\ref{fig:3_job_updating}(a) are the result of using uncapped (i.e., uncensored) data to perform the updates.  In contrast, the job updates for the realizations in Figure~\ref{fig:3_job_updating}(b) were performed by artificially introducing a power cap and censoring the historical data at its 95$\th$ percentile (i.e., the horizontal line in Figure~\ref{fig:3_job_updating}(b)).  Thus, it is unclear to the estimation procedure how large the power draw could be when it hits this threshold, and it must borrow strength from the parent model.

Although the job updates via MCMC are fast, they could possibly benefit from a sequential Monte Carlo (SMC) (a.k.a particle filtering) approach \citep{Liu98,Pitt99,Doucet01,Del06}. However, there are many fixed (over time) parameters in the job power model, e.g., $\pi_{j,k}, \mu_{j,k}$.  The fact that these parameters are fixed plays a critical role in the model because a job typically reverts back to its previous regimes.  SMC methods can be very challenging to apply in the presence of fixed parameters \citep{Liu01}.  Some recent advances have been made in this area however \citep{Storvik02,Andrieu05,Polson08}.  It is a subject of further work to explore an SMC alternative for job updates prior to full implementation of the proposed power capping approach.

\vspace{-.15in}
\subsection{Assessing Prediction Accuracy}
\vspace{-.05in}
\label{sec:job_CV}

\begin{figure}[h!]
\begin{center}
\vspace{-.0in}
\caption{Performance degradation (percent compute time increase) bound as a function of power cap for the three example jobs from Figure~\ref{fig:3_job_pow}.  Jobs were updated using 200 minutes of history with right censoring (i.e., power cap) at the $95\th$ percentile of the historical data.  Predictions of the performance degradation for a future time horizon of 5 minutes were obtained on a grid of potential power caps.  Predictions are summarized by the posterior mean and 95\% credible bands and compared to the \emph{actual} degradation computed from the actual power draw in the next five minutes.}
\vspace{-.1in}
\includegraphics[width=.85\textwidth]{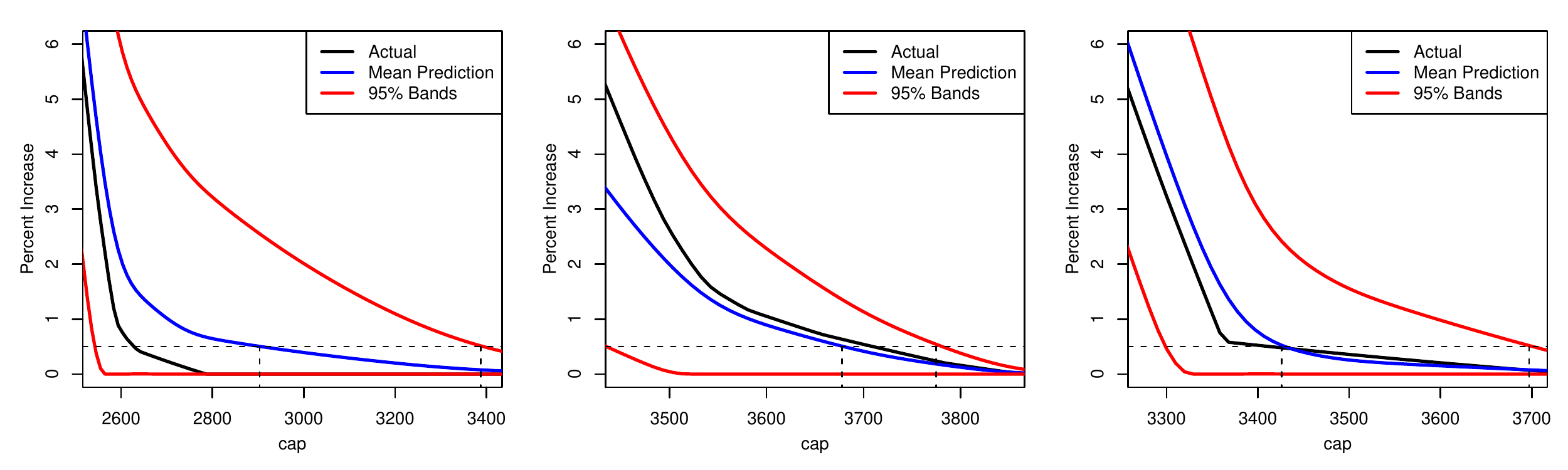}
\label{fig:3_job_pow_vs_cap}
\end{center}
\vspace{-.35in}
\end{figure}

\begin{figure}[h!]
\vspace{.1in}
\begin{center}
\caption{Normal Q-Q Plots resulting from prediction of performance degradation for each of the 213 jobs in the data set.  Several prediction scenarios are considered by varying the targeted degradation (0.5\% or 2\%), and the history, i.e., the length of the time the job had been observed (0, 30, or 200 minutes).  In each case, the historical data was assumed to be right censored by a power cap at the $95\th$ percentile of the historical data.  Simultaneous 95\% confidence bands under normality are also provided.}
\vspace{-.1in}
\includegraphics[width=.85\textwidth]{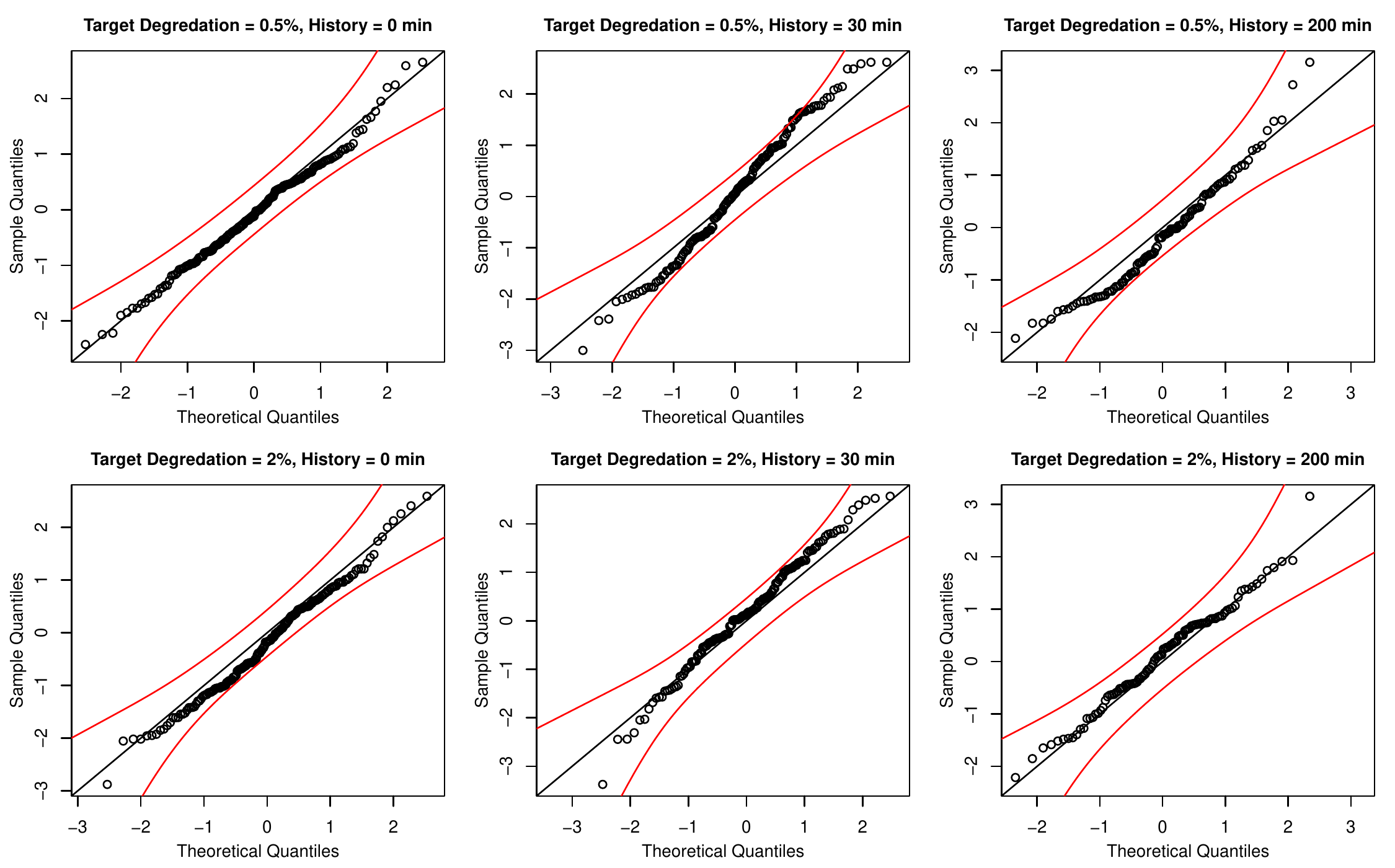}
\label{fig:QQ_plots}
\end{center}
\vspace{-.3in}
\end{figure}

The accuracy of the proposed method for the prediction of the performance degradation is assessed in Figures~\ref{fig:3_job_pow_vs_cap}~and~\ref{fig:QQ_plots}.  Figure~\ref{fig:3_job_pow_vs_cap} displays degradation bound predictions (mean and 95\% prediction bands) along with the actual degradation bound as a function of the node power cap for the three jobs displayed in Figures~\ref{fig:3_job_pow}~and~\ref{fig:3_job_updating}.  The model was updated after 200 minutes of history in each case, and then predictions of the performance degradation for the next five minutes across a range of power caps were obtained.
%In principle, a job will run a little slower when a power cap is introduced, but the procedure used here implicitly assumes that a job was running at its uncapped (max) efficiency in its previous history.  This phenomenon would need to be accounted for in the rates of transitions out of regimes in the model if severe power caps were to be introduced, i.e., caps leading to significant performance degradations.  However, the goal of this work is to select a power cap such that minimal performance degradation occurs, i.e., at most in the $\sim 5$\% range.  In such cases, this small bias in regime transtion rates leads to negligible differences in prediction of degradation (particularly for a short time horizon).

The prediction accuracy across all 213 jobs in the dataset can be assessed via the Q-Q plots in Figure~\ref{fig:QQ_plots}.  The actual degradation is compared to the predictive distribution for a targeted degradation (0.5\% or 2.0\%) based on the mean prediction.  The actual degradation for each job is converted to a Z-score via the respective predictive distribution.  A normal Q-Q plot is then produced for the various prediction scenarios of the amount of history used for updating (0, 30, or 200 minutes) and target degradation (0.5\% or 2.0\%).  For the purpose of updating a job, the historical data for each job was assumed to be right censored by a power cap at the $95\th$ percentile of the historical data.
Simultaneous 95\% confidence bands for each Q-Q plot were computed under the assumption that the model is correct and are displayed for reference.  In all cases, the Z-scores created from the predictive distribution fall inside the confidence bands, indicating that there is little to no sign of model inadequacy.

\vspace{-.15in}
\subsection{A Pragmatic Alternative to the Bayesian Model}
\vspace{-.05in}
\label{sec:freq_estimator}

An alternative to the full Bayesian model and MCMC is to use a simple, pragmatic approach to estimate the parameters of the model in (\ref{eq:job_pow}) for each job.  For example, the $\mu_{j,k}$ for the $j\th$ job can be estimated using normal mixture clustering such as that in \cite{Fraley02}, provided by the \verb1mclust1 package in R.  Conditional on the $\mu_{j,k}$, the value of the hidden $\xi_j$ process could be simply inferred based on the most likely group membership from the normal mixture model.  Once the $\mu_{j,k}$ and $\xi_j$ are assumed known, the parameters of the time to transition, transition probabilities, the AR(1) process and the observation error can be estimated via maximum likelihood estimates (MLEs).  Let the estimate of all these parameters for the $j\th$ job be denoted $\hTheta_j$.  The parent distribution for each of the parameters could be taken to be the empirical distribution formed by the collection of $\{\hTheta_j\}$.  The major advantage of this approach over the fully Bayesian approach is that it is incredibly simple and fast to implement.  While it assumes the same model in (\ref{eq:job_pow}), it makes far fewer assumptions regarding prior distributions.  However, the major disadvantage is that it ignores the uncertainty in job specific parameter estimates and does not penalize toward the parent distribution to borrow strength when estimating the parameters of a job that has a small number of observations.

A job may be updated in such a framework by calculating the conditional distribution of its parameters $\Theta^*$ given the new job's data $X^*$.  This would involve calculating the likelihood ${\cal L}(\hTheta_j)$ of the new job's data for each parameter setting in the support of the parent distribution $\{\hTheta_j\}_{j=1}^J$. The conditional distribution would be given by
\vspace{-.12in}\beq
[\Theta^* \mid X^*] \propto {\cal L}(\hTheta_j) I_{\{\Theta^* = \hTheta_j\}}.
\label{eq:freq_update}\vspace{-.12in}\eeq
The disadvantage apparent in (\ref{eq:freq_update}) is that the parent distribution was assumed to be the empirical (discrete) distribution of the $\hTheta_j$ resulting from the training data.  If the training set contains enough data so that all jobs that will run on the machine in the future will be very similar to those seen in training, then this approach will work well.  However, it will always be possible  for the machine to see entirely new jobs.  Still, this approach has the advantage of being free of many other assumptions about the parent distribution that have been made in the proposed Bayesian approach.  The performance of this approach for the purpose of node level power capping is compared to that of the full Bayesian approach next in Section~\ref{sec:analysis}.

%The same parent distributions in column 2 of Table~\ref{fig:parent_location} could be assumed, but the parameters could simply be estimated via maximum likelihood, by assuming that the realization of each job specific parameter is known to be equal to $\htheta_j$.  The major advantage of this approach is that it is incredibly simple and fast to implement.  It also makes fewer assumptions about prior distributions.  One could even go a step further and treat the 

%%%%%%%%%%%%%%%%%%%%%%%%%%%%%%%%%%%%%%%%%%%%%%%%%%%%%%%%%%%%%%%%%%%%%%%%%%%%%%%%%%%%%

\vspace{-.2in}
\section{Optimal Power Capping Across an Entire Machine}
\vspace{-.1in}
\label{sec:analysis}

In this section, a node-level power capping strategy is proposed and evaluated on a simulation study on the hypothetical Sol machine.  The simulation setup assumes that all jobs on Sol encompass multiples of 10 nodes (i.e., a cage), i.e., since only cage level data was available.  The 213 distinct Luna jobs in the data set were resampled and ``launched'' on the Sol machine in the sampled order until the machine no longer had room for the next job.  Each job was required to cover the same number of cages as it did in reality on Luna.  For example, if Sol had two cages idle, but the next job in the queue required three cages, it would have to wait until another job finished before it could start.  Jobs finished after running the same amount of time as they really did on Luna.  Once a job finished, if there was then enough room for the next job in the queue, then it was launched at that time.  If more than one job from the top of the queue would fit, then all such jobs were added.  This process was run out to steady state ($\sim 1000$ completed jobs).  This queuing strategy is far simpler than the actual queuing system used at LANL, but it is only intended to provide a realistic job mix with which to test the capping strategies.  At steady state, for example, $\sim 100$ jobs will be running with varying start times and consequently a varying amount of time history.

For a given job mix at steady state, the following scenario is considered. Cage level caps  must be imposed so that the entire system is subject to a power cap of 575~kW.  With a 56.5~kW baseline, this means the sum of the node caps (or cage caps in this case) must be 518.5~kW.  All idle cages automatically receive a fixed cap of 1.2~kW (i.e., barely above idle power draw).  All cages running the same job receive the same cap.  Therefore, for simplicity, consider the cap vector to be optimized as $\bc = [c_1,c_2,\dots,c_{J^*}]'$ containing the caps for each of the $J^*$ running jobs.  Since depending on the job mix, a number ($N_{\mbox{\scriptsize idle}}$) of cages may be at idle, the constraint becomes $\sum_j c_j \leq 518.5-1.2 N_{\mbox{\scriptsize idle}}$~kW.
We consider three power capping strategies all based on the predicted performance degradation (bound) for the next five minutes:
\begin{itemize}
\item[(i)] Minimize the weighted mean performance degradation.  That is, find the cage level power cap vector $\bc_{\mbox \scriptsize avg}$, where
\vspace{-.12in}  \bdm
  \bc_{\mbox \scriptsize avg} = \arg\min_{\bc} \left\{ \frac{\sum_j N_j \E [D_j]}{\sum_j N_j} \right\},
\vspace{-.07in}  \edm
  where $D_j$ is the performance degradation for the $j\th$ job in the next five minutes, and $N_j$ is the number of cages used by the $j\th$ job.
\item[(ii)] Minimize the expected maximum performance degradation, i.e., find the cage level power cap vector $\bc_{\mbox \scriptsize max}$, where
\vspace{-.12in}  \bdm
  \bc_{\mbox \scriptsize max} = \arg\min_{\bc}\left\{  \E \left[\max_j D_j \right] \right\}
\vspace{-.12in}  \edm
\item[(iii)] Set each cage running a job to have the same power cap,
\vspace{-.12in}  \bdm
  \bc_{\mbox \scriptsize naive} = \frac{518.5-1.2 N_{\mbox{\scriptsize idle}}}{154-N_{\mbox{\scriptsize idle}}}.
\vspace{-.12in}  \edm
\end{itemize}

The first two strategies above were applied using both the fully Bayesian approach  for estimation and updating and the simple pragmatic approach described in Section~\ref{sec:freq_estimator}.  In each case, 1000 realizations of the future job power draw were generated for each job.  And then the \verb1optim1 function in R with the Nelder-Mead algorithm was used to find the optimal cap vector for each of the two criteria.  To enforce the sum to $518.5-N_{\mbox{\scriptsize idle}} 1.2$~kW constraint, the cap vector $\bc$ was reparameterized to a vector $\bc^*$ where the value of the first element of $\bc^*$ was fixed and the remaining $J^*-1$ were allowed to vary freely.  The mapping back to $\bc$ is
\vspace{-.05in}\bdm
\bc = \bc^* \left( \frac{518.5-1.2N_{\mbox{\scriptsize idle}}}{\sum_j c_j^*} \right).
\vspace{-.05in}\edm
%Complete freedom for the other values could result in pathological/unrealistic choices for caps, so an exponential penalty was simply imposed in the optimization to discourage departure from reasonable limits, i.e.,  1.2~kW$\:<c_j<\:$4.5~kW.  Once this penalty was in place, all solutions were well inside of this range anyhow.

\begin{figure}[t]
\vspace{-.11in}
\begin{center}
\caption{Actual degradation results for the weighted average increase and the maximum increase, respectively, according to the five capping strategies. Boxplots are created from the degradation's that would have occurred for each of the 100 job mix realizations.}
\vspace{-.15in}
\includegraphics[width=.40\textwidth]{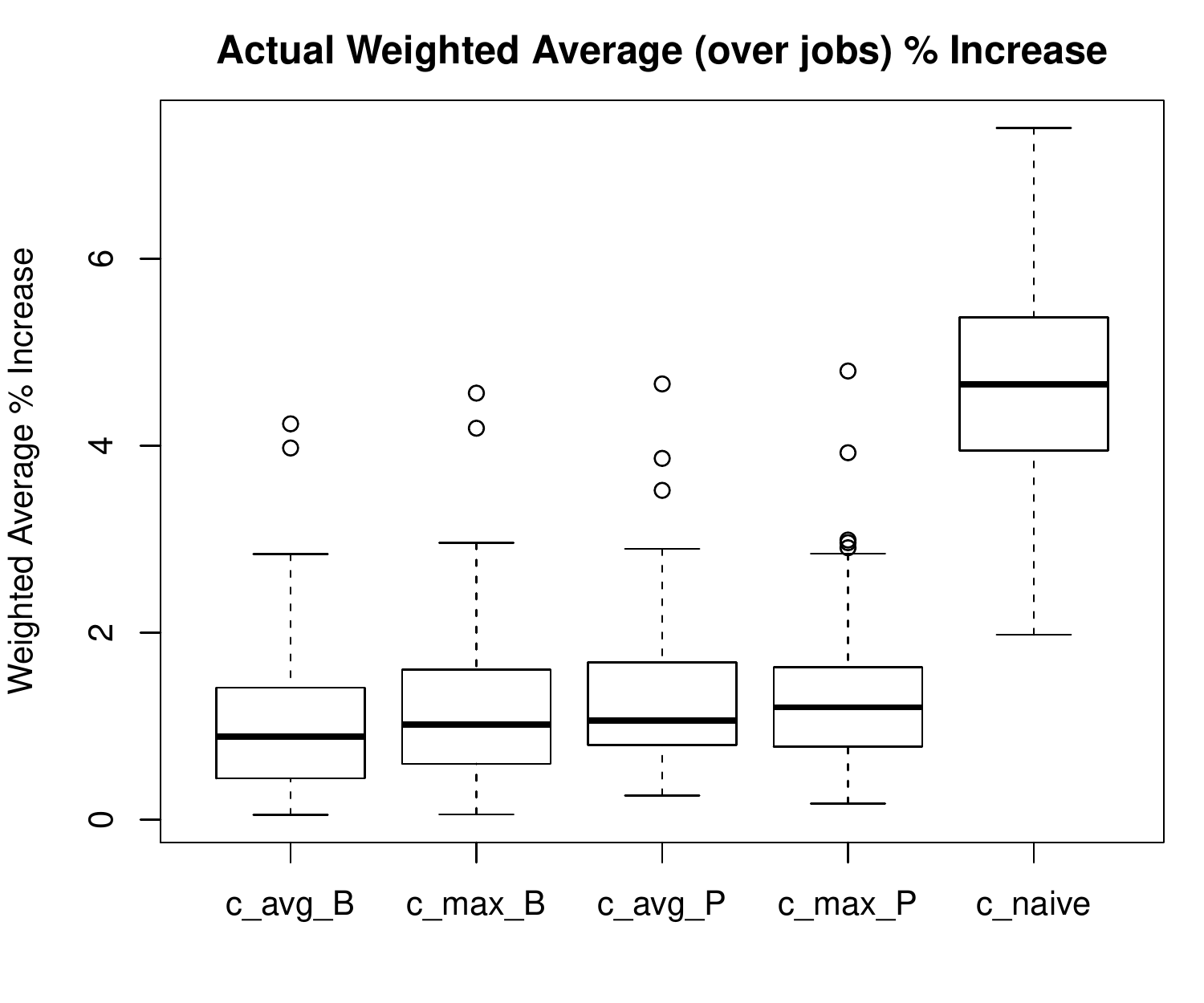}
\includegraphics[width=.40\textwidth]{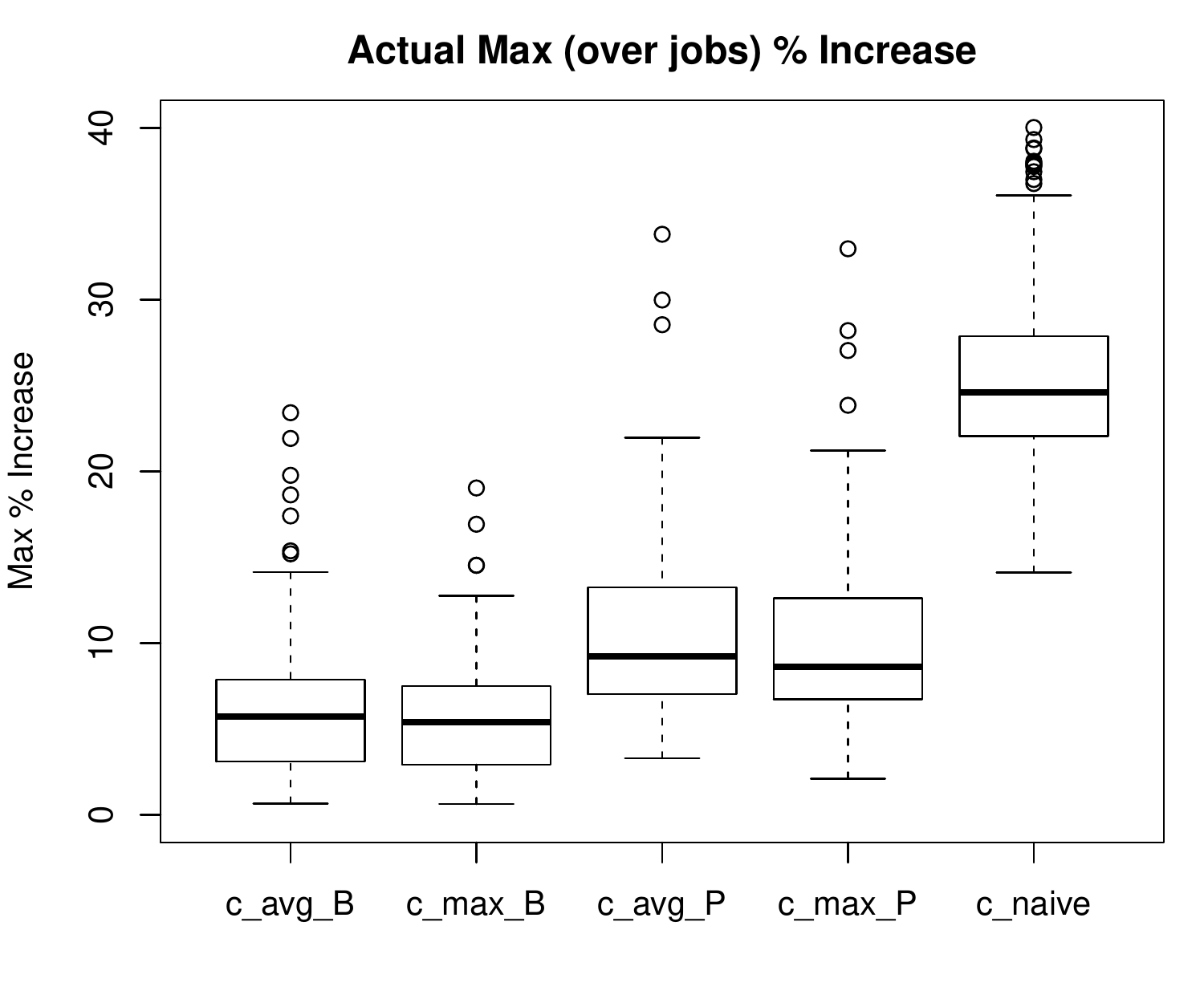}
\label{fig:boxplots}
\end{center}
\vspace{-.4in}
\end{figure}

Five capping strategies were considered, (i) \verb1c_avg_B1 : $\bc_{\mbox \scriptsize avg}$ using the full Bayesian approach, (ii) \verb1c_max_B1 : $\bc_{\mbox \scriptsize max}$ using the full Bayesian approach, (iii) \verb1c_avg_P1 : $\bc_{\mbox \scriptsize avg}$ using the pragmatic estimation approach in Section~\ref{sec:freq_estimator}, (iv) \verb1c_max_P1 : $\bc_{\mbox \scriptsize max}$ using the pragmatic estimation approach in Section~\ref{sec:freq_estimator}, (v) \verb1c_naive1 : $\bc_{\mbox \scriptsize naive}$.  These five strategies were applied to the Sol machine once it had reached a steady state setting for job scheduling on 100 different randomly generated job mixes.  The actual job performance degradation (bound) for the next five minutes was then calculated for each job and two summaries were calculated for each capping strategy: (i) the actual weighted average degradation over all running jobs and (ii) the actual maximum performance degradation across all running jobs.   Figure~\ref{fig:boxplots} displays the resulting box plots of these two metrics from the 100 simulated scenarios for each of the five capping strategies.

It is clear from Figure~\ref{fig:boxplots} that all of the proposed statistical capping strategies far outperform the naive (same cap for each job) approach.  The naive approach results in a weighted average increase of $\sim 4$\% (on average over the 100 job mix realizations), whereas the weighted average increase for the statistical strategies is $\sim 1$\% on average.  To put this in perspective, this 3\% advantage of computational efficiency would be the equivalent of freeing up 4.5 cages (i.e., 45 nodes) on Sol for additional computation.  The overhead required to use this approach would be $< 2$ nodes for the full Bayesian approach and $< 1$ node for the pragmatic estimation approach.

The naive approach also produces a max increase of $\sim 24$\% on average over the 100 realizations and as much as a 40\% increase in some cases.  In contrast, the Bayesian statistical strategies keep the maximum increase at about 6\% on average.   As would be expected, \verb1c_avg_B1 and \verb1c_avg_P1 perform better than \verb1c_max_B1 and \verb1c_max_P1 on minimizing the weighted average increase.  However, the opposite is true when using the observed maximum increase as the performance metric.  All of these differences are statistically significant. %, even if not all that important, practically.

The results for the pragmatic approach are very competitive with the full Bayesian approach at minimizing the weighted average increase.  The full Bayesian approach, however, does a better job of minimizing the maximum job increase.  This makes some intuitive sense, as accurate prediction of the expected maximum would rely more heavily on representation of uncertainty.  And the Bayesian approach addresses much of the estimation uncertainty that the pragmatic approach inherently ignores.  Still the pragmatic approach is an order of magnitude faster than the full Bayesian approach, and may remain in the discussion for implementation purposes, depending on the performance goal.  In either case, there is always the potential of model failure for pathological jobs that are unlike any in the training data.  While this is certainly a concern, the only real danger is that the efficiency gain illustrated in Figure~\ref{fig:boxplots} is reduced during the execution of such a job.  Since these jobs are uncommon by definition, this should have little effect on overall performance.

%%%%%%%%%%%%%%%%%%%%%%%%%%%%%%%%%%%%%%%%%%%%%%%%%%%%%%%%%%%%%%%%%%%%%%%%%%%%%%%%%%%%%

\vspace{-.2in}
\section{Conclusions \& Further Work}
\vspace{-.1in}
\label{sec:conclusions}

A novel statistical model has been developed for the power used by a HPC jobs.  To the best of our knowledge, this is the first attempt to statistically model this process.  This model was then used to inform an intelligent node-level power capping strategy, with the intention that this could be used for systems in the near future that have a node-level capping mechanism.  This approach has been demonstrated via a simulation molded from a real machine at LANL.  The results demonstrate that the proposed approach is about $5 \times$ more efficient than the simple approach where all nodes receive the same power cap.  In addition, the job power model introduced here could have applications beyond power capping, such as intelligent scheduling, optimizing power contracts with utilities, improving the power efficiency of jobs, etc.  

There are two important areas where this approach could be further refined.  In particular,  the current model assumes that new jobs come from a large population of all jobs that could run on the machine.  This results in a large amount of uncertainty when predicting the future of a new job prior to seeing any data from it.
%While, it would be difficult to make use of a job label (as these are not that informative and change with updates to code, etc.),
Thus, it may be beneficial to introduce another \emph{user} level into the hierarchical model.  In other words, a new job could come from a specific user's population, as opposed to the population of all possible jobs.  As mentioned previously, the MCMC approach to update jobs is currently fast enough for practical application.  However, it would still be prudent to explore possible SMC solutions to job updating in order to further increase the computational efficiency.  Finally, results of Section~\ref{sec:analysis} did not consider queuing strategy at all.  The queuing system used at LANL has a very complicated set of rules for priority, etc.  However, it would be interesting to consider queuing and capping strategies simultaneously within some realistic constraints to achieve the most efficient end result.

{\singlespacing
\small
  \bibliography{curt_ref.bib}
\bibliographystyle{agsm}
}

\clearpage

\begin{appendix}

\setcounter{equation}{0}
\setcounter{page}{1}
\renewcommand{\theequation}{A\arabic{equation}}

\vspace{-.2in}
\begin{Large}
{\bf
  Supplementary Material: ``Modeling and Predicting Power Consumption of High Performance Computing Jobs''}
\end{Large}
\vspace{-.3in}

\section{MCMC Computational Details}
\vspace{-.05in}
\label{sec:computation}

As mentioned in the main paper, each iteration of the MCMC algorithm for parent parameter estimation consists of (i) updating the job specific parameters $\Theta_j$, $j=1,\dots,J$, and (ii) updating the parent parameters $\Theta^P$.  The MCMC algorithm for updating a new job given its data up until the current time consists of just iterating over (i) for that single new job.  The full conditionals for Gibbs sampling and/or MH steps used are provided in Section~\ref{sec:job_updates} for the elements of $\Theta_j$, and in Section~\ref{sec:parent_updates} for the elements of $\Theta^P$.

\subsection{Updates for Job Specific Parameters}
\label{sec:job_updates}

The entire collection of parameters to be sampled for the $j\th$ job is
\bdm
\Theta_j = \left\{
\left\{\xi_j(t)\right\}_{t=1}^{T_j},
\left\{\lambda_{j,k}\right\}_{k=1}^K,
\left\{\pi_{j,k}\right\}_{k=1}^K,
\left\{\mu_{j,k}\right\}_{k=1}^K,
\left\{z_j(t)\right\}_{t=1}^{T_j},
\sigma_j^2,
\rho_j
\right\}, \; j=1,\dots,J.
\label{eq:job_param_set}
\edm
As mentioned in the main paper, the number of components in stick-breaking model for $\pi_{j,k}$ was capped at a finite value $K$, for computational convenience.  The value of $\pi_{j,K}$ was observed and $K$ was increased until $\pi_{j,K}$ values were negligible for all jobs at $K=10$.  Before proceeding to the full conditionals for the elements of $\Theta_j$ the following equivalent model for the $\xi_j(t)$ is introduced because it allows conjugate updates for the $\pi_{j,k}$.

Assume that the time to a \emph{possible} transition when in regime $\xi_j(t)=k$ is
\beq
T^* \sim \mbox{Geometric}\left(\lambda_{j,k}\right).
\label{eq:trans_time_equiv}
\eeq
At a possible transition time $T^*=u$ a regime label is chosen according to
\beq
\Pr(\xi_j(u)=l \mid \xi_j(u-1)=k) = \pi_{j,l},  \;\; \mbox{for} \; l=1,\dots,K 
\label{eq:xi_transition_probs_equiv}
\eeq
That is, at the time of a possible transition out of state $k$, a regime $l$ is chosen with probability equal to $\pi_{j,l}$.  The difference here is the possibility that $l=k$ (i.e., there is possibly no regime change at time $u$).

\begin{prop}
  The model for $\xi_j(t)$ proposed in (\ref{eq:trans_time_equiv}) and (\ref{eq:xi_transition_probs_equiv}) leads to
  %transition times $T$ out of regime $k$ that are Geometric$(\lambda_{j,k})$ and transition probabilities (to a new regime $l \neq k$) that are proportional to $\pi_{j,l}$, i.e.,
  an equivalent model to that described in (\ref{eq:trans_time}) and (\ref{eq:xi_transition_probs}) in the main paper.
\end{prop}

%\proof
It is relatively straight-forward to justify Proposition~1 by recognizing that the model for $\xi_j(t)$ described in described in (\ref{eq:trans_time}) and (\ref{eq:xi_transition_probs}) is a discrete time MC, so too is that described in (\ref{eq:trans_time_equiv}) and (\ref{eq:xi_transition_probs_equiv}), and they have the same probability transition matrix.

\begin{comment}
The probability that the time $T$ until transition out of regime $k$ is greater than $n$ is the same as the probability that there are $n$ consecutive transitions from state $k$ to state $k$ in the discrete time Markov chain above.  That is,
\bdm
\Pr(T>n) = (P_{k,k})^n = \left(\lambda_{j,k}\pi_{j,k} + 1- \lambda_{j,k} \right)^n
%= \left(\frac{(1-\lambda_{j,k})(1-\pi_{j,k})}{(1-\pi_{j,k})} \right)^t
= \left[1-\lambda_{j,k}(1-\pi_{j,k}) \right]^n,
\edm
i.e., $T \sim \mbox{Geometric}(\lambda_{j,k} (1-\pi_{j,k}))$ as in (\ref{eq:trans_time}).  Conditional on a transition out of state $k$ at time $t+1$, the probability that the transition is into state $l$ is,
\bdm
\Pr\left(\xi_j(t+1) = l \;\mid\; \xi_j(t) = k \;,\; \xi_j(t+1) \neq k \right) = \frac{P_{k,l}}{1-P_{k,k}} \propto \pi_{j,l},
\edm
just as in (\ref{eq:xi_transition_probs}).
\begin{flushright}\vspace{-.425in} $\Box$ \end{flushright}
\end{comment}

With this new representation for $\xi_j$, we also introduce a new latent variable $\phi_j(t)$ to be sampled in the MCMC;  $\phi_j(t)$ is the indicator of whether or not time $t$ was a \emph{possible} transition time.  This is done to allow for conjugate updates of $\lambda_{j,k}$, conditional on $\phi_j$.  A latent variable $\psi_{j,k} \in \{1,\dots,M\}$ is also introduced, representing the index of the component in the normal mixture in (10) that produced $\mu_{j,k}$.  This is to allow for conjugate updates of the $\mu_{j,k}$ and parent parameters of the normal mixture.  The complete collection of parameters to be sampled for the $j\th$ job is then
\bdm
\Theta_j = \left\{
\left\{\xi_j(t)\right\}_{t=1}^{T_j},
\left\{\phi_j(t)\right\}_{t=1}^{T_j},
\left\{\lambda_{j,k}\right\}_{k=1}^K,
\left\{\pi_{j,k}\right\}_{k=1}^K,
\left\{\mu_{j,k}\right\}_{k=1}^K,
\left\{\psi_{j,k}\right\}_{k=1}^K,
\left\{z_j(t)\right\}_{t=1}^{T_j},
\sigma_j^2,
\rho_j
\right\}.
\label{eq:job_param_set_2}
\edm

%\clearpage
\noindent
{$\underline{\xi_j(t), \phi_j(t) \mid \mbox{rest}}$}

\noindent
Let \emph{rest} denote the data for the $j\th$ job and all parameters in $\Theta^P$ and in $\Theta_j$ except $\xi_j(t)$ and $\phi_j(t)$.
Because of the Markov property of $\xi_j(t)$, conditional on all other parameters and the data (i.e., rest), $\{\bxi_j(t), \phi_j(t)\}$ only depends on $\{\xi_j(s): s\neq t\}$ through $\xi_j(t-1)$ (for $t>1$) and $\xi_j(t+1)$ (for $t<T_j$).  That is,
\begin{eqnarray}
\Pr\left(\xi_j(t) = k \mid \mbox{rest} \right) & \! \propto \! &
\Pr\left(\xi_j(t) = k \mid \xi_j(t-1)\right)
\Pr\left(\xi_j(t+1) \mid \xi_j(t) = k\right)
{\cal L}\left(x_j(t) - \mu_{j,k} - Z_j(t)\right) \nonumber \\
& \! = \! &
P_{\xi_j(t-1),k}P_{k,\xi_j(t+1)} N(x_j(t) - \mu_{j,k} - Z_j(t);0,\tau^2),
\label{eq:xi_update}
\end{eqnarray}
where $P_{k,l}$ was defined in (\ref{eq:xi_TPM}) and $N(\cdot,0,\tau^2)$ is the Gaussian density with mean 0 and variance $\tau^2$.  Once $\xi_j(t)$ is updated via (\ref{eq:xi_update}) the indicator $\phi_j(t)$ can be updated conditional on the rest {\bf and} $\xi_j(t)$ as
\bdm
\Pr\left(\phi_j(t) = 1 \mid \mbox{rest}\;,\;\xi_j(t)\right) = \left\{
\begin{array}{ll}
  1 & \mbox{if $\xi_j(t-1) \neq \xi_j(t)$} \\
  \frac{\lambda_{j,k} \pi_{j,k}}{1-\lambda_{j,k}(1-\pi_{j,k})}  & \mbox{if $\xi_j(t-1) = \xi_j(t) = k$}.
\end{array}
\right.
\edm
\\[-.1in]

%%%%%%%%%%%%%%%%%%%%%%%%%%%%%%%%%%%%%%%%%%%%%%%%%%%%%%%%%%%%%%%%%%%%%%%%%%%%%%

\noindent
{$\underline{\lambda_{j,k} \mid \mbox{rest}} $}

\noindent
Conditional on $\Theta^P$ all other parameters in $\Theta_j$, $\lambda_{j,k}$ only depends on $\left\{\phi_j(t)\right\}_{t=1}^{T_j}$.  Specifically,
\bdm
\lambda_{j,k} \mid \mbox{rest} \sim \mbox{Beta}(a^*, b^*), 
\edm
where $a^* = \alpha_\lambda + M_{j,k}$, $b^*  = \beta_\lambda + N_{j,k} - M_{j,k}$, $M_{j,k}$ is the number of \emph{possible} transitions generated from state $k$, and $N_{j,k}$ is the number of total time steps observed from state $k$, i.e., 
\begin{eqnarray}
M_{j,k} & \!=\!& \sum_{t=1}^{T_j-1} \phi_j(t+1) I_{\{\xi_j(t)=k\}} \nonumber \\
N_{j,k} & \!= \!& \sum_{t=1}^{T_j-1} I_{\{\xi_j(t)=k\}} \nonumber
\end{eqnarray}
\\[-.1in]

%%%%%%%%%%%%%%%%%%%%%%%%%%%%%%%%%%%%%%%%%%%%%%%%%%%%%%%%%%%%%%%%%%%%%%%%%%%%%%

\noindent
{$\underline{\bpi_{j}=[\pi_{j,1},\dots,\pi_{j,K}]' \mid \mbox{rest}}$}

\noindent
There is a one to one correspondence between $\bpi_{j}$ and $\bv_j = [v_{j,1},\dots,v_{j,K}]'$ in (\ref{eq:v_jk}).  Conditional on $\Theta^P$ and the rest of $\Theta_j$, $\bv_j$ depends only on $\left\{\phi_j(t)\right\}_{t=1}^{T_j}$, $\left\{\xi_j(t)\right\}_{t=1}^{T_j}$, and $\tdelta$.  Specifically,
\bdm
v_{j,k} \mid \mbox{rest} \stackrel{ind}{\sim} \mbox{Beta}(a^*_k, b^*_k),
\edm
where,
\begin{eqnarray}
a^*_k & \!=\!& \sum_{t=1}^{T_j} \phi_j(t) I_{\{\xi_j(t)=k\}} +1 \nonumber \\
b^*_k & \!= \!& \sum_{t=1}^{T_j}\phi_j(t) I_{\{\xi_j(t)>k\}} + \tdelta \nonumber
\end{eqnarray}
\\[-.1in]

%%%%%%%%%%%%%%%%%%%%%%%%%%%%%%%%%%%%%%%%%%%%%%%%%%%%%%%%%%%%%%%%%%%%%%%%%%%%%%

\noindent
{$\underline{\mu_{j,k} \mid \mbox{rest}}$}

\noindent
Let $\bx_j^{(k)}$ be the vector (ordered in time) of all $x_j(t)$ when $\xi_j(t)=k$.  Define the vectors $\bz_j^{(k)}$ and $\beps_j^{(k)}$ analogously.  Then define
$\br_j^{(k)} = \bx_j^{(k)} - \bz_j^{(k)} = \mu_{j,k} - \beps_j^{(k)}$, and let $n_{j,k}$ be the length of the $\br_j^{(k)}$ vector.  Lastly for convenience of notation, let the current value of $\psi_{j,k}$ be denoted as $m$.  All \emph{observations} $\br_j^{(k)}$ are \emph{iid} from the same normal distribution with known variance and mean $\mu_{j,k} \sim N(\nu_m, \varsigma^2_m)$, resulting in a simple conjugate update, i.e.,
\bdm
\mu_{j,k} \mid \mbox{rest} \sim N(\mu^*, {\sigma^2}^*),
\edm
where,
\begin{eqnarray}
  \mu^* & \!=\!& {\sigma^2}^* \left( \frac{\nu_m}{\varsigma^2_m} + \frac{\sum_i r_{j,i}}{\tau^2} \right)  \nonumber \\
{\sigma^2}^* & \!= \!& \left(\frac{1}{\varsigma^2_m} + \frac{\sum_i n_{j,k}}{\tau^2} \right)^{-1}.   \nonumber
\end{eqnarray}
\\[-.1in]

%%%%%%%%%%%%%%%%%%%%%%%%%%%%%%%%%%%%%%%%%%%%%%%%%%%%%%%%%%%%%%%%%%%%%%%%%%%%%%

\noindent
{$\underline{\psi_{j,k} \mid \mbox{rest}}$}

\noindent
$\Pr(\psi_{j,k} = m \mid \mbox{rest} ) \propto \tomega_m N(\mu_{j,k}; \tnu_m, \tvarsigma^2_m)$ \\[-.1in]

%%%%%%%%%%%%%%%%%%%%%%%%%%%%%%%%%%%%%%%%%%%%%%%%%%%%%%%%%%%%%%%%%%%%%%%%%%%%%%

\noindent
{$\underline{z_j(t) \mid \mbox{rest}}$}

\noindent
Let $r_{j}(t) = x_j(t) - \mu_{j,\xi_j(t)} = z_j(t) + \eps_j(t)$, and let the time ordered vector of the $r_{j}(t)$ be denoted $\br_j$. All \emph{observations} $\br_j$ are \emph{ind} from a normal distribution with known variance and mean vector $\bz_j \sim N(\bzero, \sigma^2 \bGamma_{\rho_j})$, where $\bGamma_{\rho_j}$ is correlation matrix for $\bz_j$ formed by evaluating the correlation function in (\ref{eq:zj_model}) at the observed time points for the $j\th$ job.  
\bdm
\bz_{j} \mid \mbox{rest} \sim N(\bmu^*, {\bSigma}^*),
\edm
where,
\begin{eqnarray}
  \bmu^* & \!=\!& {\bSigma}^* \br_j  \nonumber \\
      {\bSigma}^* & \!= \!& \tau^2\left( \bI + \frac{\tau^2}{\sigma^2_j} \bGamma_{\rho_j} \right)^{-1}.
      \label{eq:Sigma_z}
\end{eqnarray}
It is far more efficient to use the kalman filter or use Gaussian Markov random fields (GMRF) results \citep{Rue05} in this case as opposed to actually evaluating the inverse in (\ref{eq:Sigma_z}).  We use the latter here, let $\bQ = 1/\sigma_j^2 \bGamma_{\rho_j}^{-1}$, i.e., the precision matrix for the prior on $\bz_j$.  This matrix is readily obtainable without a matrix decomposition due the Markov model imposed by the exponential covariance function.  The precision matrix $({\bSigma}^*)^{-1}$ for the update of $\bz_j$ is a sparse matrix with a bandwidth of 3.  Efficient algorithms exist for generating a multivariate normal vector in such cases, see page 31 of \cite{Rue05} for example.\\[-.1in]

%%%%%%%%%%%%%%%%%%%%%%%%%%%%%%%%%%%%%%%%%%%%%%%%%%%%%%%%%%%%%%%%%%%%%%%%%%%%%%

\noindent
{\underline{MH update for $\sigma_j^2$}}

\noindent
As mentioned in the main paper, the full conditional distribution of $\sigma_j^2 \mid \mbox{rest}$ does not have a convenient form with which to perform Gibbs updates.  However, the MH ratio has a simple form which is easy to compute.  The prior used is $\sigma_j^2 \sim \log N(\tmu_\sigma, \tsigma^2_\sigma)$.  Proposals for ${\sigma_j^2}^*$ are made by identifying an $a$ and $b$ for an IG$(a,b)$ distribution with the same mean and variance as the log-normal prior.  By assuming the prior for $\sigma_j^2 \sim \mbox{IG}(a,b)$, the update is conjugate.  Thus we take the proposal to be this conjugate update, specifically,
\bdm
{\sigma_j^2}^* \sim \mbox{IG}(a + T_j/2\;,\; b+\sum_t Z_j(t)^2).
\edm
That is, the proposals are independent of the current $\sigma_j^2$ value, let the density of the proposal be denoted $d({\sigma_j^2}^*)$.  The only portion of the full model likelihood that differs between the current value and the proposal is ${\cal L}(\bz_j; \sigma_j^2, \rho_j)$ which is a multivariate normal density with mean 0 and covariance $\sigma_j^2 \bGamma_{\rho_j}$.  As with the update of $\bz_j$, there are efficient means of evaluating this density (or the log-density) via GMRF results.  The MH ratio is then
\bdm
MH = \frac{\cL(\bz_j; {\sigma_j^2}^*, {\rho_j}) \pi({\sigma_j^2}^*) d(\sigma_j^2)}
{\cL(\bz_j; {\sigma_j^2}, \rho_j) \pi({\sigma_j^2}) d({\sigma_j^2}^*)}.
\edm
\\[-.1in]

%%%%%%%%%%%%%%%%%%%%%%%%%%%%%%%%%%%%%%%%%%%%%%%%%%%%%%%%%%%%%%%%%%%%%%%%%%%%%%

\clearpage
\noindent
{\underline{MH update for $\rho_j$}}

\noindent
As mentioned in the main paper, the $\rho_j$ were updated via a MH random walk proposals.  However, the random walk was conducted on the log scale, i.e., $\log(\rho_j^*) = \log(\rho_j+\epsilon)$ for a deviate $\epsilon \sim N(0,s^2)$.  With the use of the log scale, a constant tuning parameter $s^2=0.25$ could be used for all jobs to achieve acceptances across all jobs in the range of (30\% - 55\%).  Let the density of the proposal, given the current value of $\rho_j$ be denoted $d({\rho_j}^* \mid {\rho_j})$.  As with updates of $\sigma_j^2$ above, the only portion of the full model likelihood that differs between the current value and the proposal is ${\cal L}(\bz_j; \sigma_j^2, \rho_j)$.   The MH ratio is then
\bdm
MH = \frac{\cL(\bz_j; \sigma_j^2, \rho_j^*) \pi(\rho_j^*) d(\rho_j \mid \rho_j^*)}
{\cL(\bz_j; \sigma^2, \rho_j) \pi({\rho_j}) d(\rho_j^* \mid \rho_j)}.
\edm
\\[-.1in]

%%%%%%%%%%%%%%%%%%%%%%%%%%%%%%%%%%%%%%%%%%%%%%%%%%%%%%%%%%%%%%%%%%%%%%%%%%%%%%

%%%%%%%%%%%%%%%%%%%%%%%%%%%%%%%%%%%%%%%%%%%%%%%%%%%%%%%%%%%%%%%%%%%%%%%%%%%%%%

%%%%%%%%%%%%%%%%%%%%%%%%%%%%%%%%%%%%%%%%%%%%%%%%%%%%%%%%%%%%%%%%%%%%%%%%%%%%%%

%%%%%%%%%%%%%%%%%%%%%%%%%%%%%%%%%%%%%%%%%%%%%%%%%%%%%%%%%%%%%%%%%%%%%%%%%%%%%%

%%%%%%%%%%%%%%%%%%%%%%%%%%%%%%%%%%%%%%%%%%%%%%%%%%%%%%%%%%%%%%%%%%%%%%%%%%%%%%

\subsection{Updates for Parent Parameters}
\label{sec:parent_updates}

The entire collection of parent parameters sampled in the MCMC is
\bdm
\Theta^P = \left\{\tmu_\sigma, \tsigma_\sigma, \tmu_\rho, \tsigma_\rho, \left\{\tomega_m\right\}_{m=1}^M, \left\{\tnu_m\right\}_{m=1}^M, \left\{\tvarsigma^2_m\right\}_{m=1}^M, \talpha_\lambda, \tbeta_\lambda, \tgamma, \tdelta, \ttau^2 \right\}
\edm

\noindent
{$\underline{\tmu_\sigma \mid \mbox{rest}}$}

\noindent
Conditional on the rest, $\tmu_\sigma$ has a simple conjugate normal update,
\bdm
\tmu_\sigma \mid \mbox{rest} \sim N(\mu^*, {\sigma^2}^*),
\edm
where,
\begin{eqnarray}
  \mu^* & \!=\!& {\sigma^2}^* \left( \frac{M_\sigma}{S^2_\sigma} + \frac{\sum_{j=1}^J \log(\sigma^2_j)}{\tsigma_\sigma^2} \right)  \nonumber \\
{\sigma^2}^* & \!= \!& \left(\frac{1}{S^2_\sigma} + \frac{J}{\tsigma^2_\sigma} \right)^{-1}.   \nonumber
\end{eqnarray}
\\[-.1in]

%%%%%%%%%%%%%%%%%%%%%%%%%%%%%%%%%%%%%%%%%%%%%%%%%%%%%%%%%%%%%%%%%%%%%%%%%%%%%%

\noindent
{$\underline{\tsigma_\sigma \mid \mbox{rest}}$}

\noindent
Conditional on the rest, $\tsigma^2_\sigma$ has a simple conjugate IG update,
\bdm
\tsigma^2_\sigma \sim \mbox{IG}\left(\alpha_\sigma + \frac{J}{2}\;,\; \beta_\sigma + \frac{1}{2}\sum_{j=1}^J\left[\log(\sigma^2_j) - \tmu_\sigma\right]^2\right).
\edm
\\[-.1in]

%%%%%%%%%%%%%%%%%%%%%%%%%%%%%%%%%%%%%%%%%%%%%%%%%%%%%%%%%%%%%%%%%%%%%%%%%%%%%%

\noindent
{$\underline{\tmu_\rho \mid \mbox{rest}}$}

\noindent
Conditional on the rest, $\tmu_\rho$ has a simple conjugate normal update,
\bdm
\tmu_\rho \mid \mbox{rest} \sim N(\mu^*, {\sigma^2}^*),
\edm
where,
\begin{eqnarray}
  \mu^* & \!=\!& {\sigma^2}^* \left( \frac{M_\rho}{S^2_\rho} + \frac{\sum_{j=1}^J \log(\rho_j)}{\tsigma_\rho^2} \right)  \nonumber \\
{\sigma^2}^* & \!= \!& \left(\frac{1}{S^2_\rho} + \frac{J}{\tsigma^2_\rho} \right)^{-1}.   \nonumber
\end{eqnarray}
\\[-.1in]

%%%%%%%%%%%%%%%%%%%%%%%%%%%%%%%%%%%%%%%%%%%%%%%%%%%%%%%%%%%%%%%%%%%%%%%%%%%%%%

\noindent
{$\underline{\tsigma_\rho \mid \mbox{rest}}$}

\noindent
Conditional on the rest, $\tsigma^2_\rho$ has a simple conjugate IG update,
\bdm
\tsigma^2_\rho \sim \mbox{IG}\left(\alpha_\rho + \frac{J}{2} \;,\; \beta_\rho + \frac{1}{2}\sum_{j=1}^J\left[\log(\rho_j) - \tmu_\rho\right]^2\right).
\edm
\\[-.1in]

%%%%%%%%%%%%%%%%%%%%%%%%%%%%%%%%%%%%%%%%%%%%%%%%%%%%%%%%%%%%%%%%%%%%%%%%%%%%%%

%\clearpage
\noindent
{$\underline{\bomega=[\tomega_{1},\dots,\tomega_{M}]' \mid \mbox{rest}}$}

\noindent
There is a one to one correspondence between $\bomega$ and $\bu = [u_{1},\dots,u_{M}]'$ in (\ref{eq:u_m}).  Conditional on $\Theta^P$ and the rest of $\Theta_j$, $\bu$ depends only on the $\left\{\left\{\psi_{j,k}\right\}_{j=1}^{J}\right\}_{k=1}^{K}$ and $\tgamma$.  Specifically,
\bdm
u_{m} \mid \mbox{rest} \stackrel{ind}{\sim} \mbox{Beta}(a^*_m, b^*_m),
\edm
where,
\begin{eqnarray}
a^*_m & \!=\!& \sum_{j=1}^{J} \sum_{k=1}^K I_{\{\psi_{j,k}=m\}} + 1 \nonumber \\
b^*_m & \!= \!& \sum_{j=1}^{J} \sum_{k=1}^K I_{\{\psi_{j,k}>m\}} + \tgamma \nonumber
\end{eqnarray}
\\[-.1in]

%%%%%%%%%%%%%%%%%%%%%%%%%%%%%%%%%%%%%%%%%%%%%%%%%%%%%%%%%%%%%%%%%%%%%%%%%%%%%%

\noindent
{$\underline{\tnu_m \mid \mbox{rest}}$}

\noindent
Let $\bmu^{(m)}$ be the vector of all $\mu_{j,k}$ when $\psi_{j,k}=m$, and let $n_m$ denote the length of this vector.  All \emph{observations} $\bmu^{(m)}$ are \emph{iid} from the same normal distribution with known variance and mean $\tnu_{m} \sim N(M_\nu, S^2_\nu)$, resulting in a simple conjugate update, i.e.,
\bdm
\tnu_m \mid \mbox{rest} \sim N(\mu^*, {\sigma^2}^*),
\edm
where,
\begin{eqnarray}
  \mu^* & \!=\!& {\sigma^2}^* \left( \frac{M_\nu}{S^2_\nu} + \frac{\sum_{i=1}^{n_m} \mu_i^{(m)}}{\varsigma_m^2} \right)  \nonumber \\
{\sigma^2}^* & \!= \!& \left(\frac{1}{S^2_\nu} + \frac{n_m}{\varsigma^2_m} \right)^{-1}.   \nonumber
\end{eqnarray}
\\[-.1in]

%%%%%%%%%%%%%%%%%%%%%%%%%%%%%%%%%%%%%%%%%%%%%%%%%%%%%%%%%%%%%%%%%%%%%%%%%%%%%%

%\clearpage
\noindent
{$\underline{\tvarsigma^2_m \mid \mbox{rest}}$}

\noindent
Conditional on the rest, $\tvarsigma^2_m$ has a simple conjugate IG update,
\bdm
\tvarsigma^2_m \sim \mbox{IG}\left(A_\varsigma + \frac{n_m}{2}\;,\; B_\varsigma + \frac{1}{2}\sum_{i=1}^{n_m}\left(\mu_i^{(m)} - \tnu_m\right)^2\right).
\edm
\\[-.1in]

%%%%%%%%%%%%%%%%%%%%%%%%%%%%%%%%%%%%%%%%%%%%%%%%%%%%%%%%%%%%%%%%%%%%%%%%%%%%%%

\noindent
{$\underline{\tgamma \mid \mbox{rest}}$}

\noindent
Conditional on the rest, $\tgamma$ depends only on the $\tomega_{m}$ (or equivalently the $u_{m}$).   One can equivalently think of the SB model as a prior for $u_m = \Pr(\psi_{j,k} = m \mid \psi_{j,k} > m-1) \stackrel{iid}{\sim} \mbox{Beta}(1,\tgamma)$, for $m=1,\dots,M$.  A Gamma prior is conjugate for $\tgamma$ in this model, thus, $\tgamma$ has the update,
\bdm
\tgamma  \sim \mbox{Gamma}\left(A_\gamma + M \;,\;
B_\gamma - \sum_{m=1}^{M-1}\log (1-u_m) \right).
\edm
\\[-.1in]

%%%%%%%%%%%%%%%%%%%%%%%%%%%%%%%%%%%%%%%%%%%%%%%%%%%%%%%%%%%%%%%%%%%%%%%%%%%%%%

\noindent
{$\underline{\tdelta \mid \mbox{rest}}$}

\noindent
Conditional on the rest, $\tdelta$ depends only on the $\pi_{j,k}$ (or equivalently the $v_{j,k}$).  By a completely analogous argument, as that for the update of $\tgamma$ above, $\tdelta$ has a conjugate Gamma update,
\bdm
\tdelta  \sim \mbox{Gamma}\left(A_\delta + JK \;,\;
B_\delta - \sum_{j=1}^J\sum_{k=1}^{K-1}\log (1-v_{j,k}) \right).
\edm
\\[-.1in]

%%%%%%%%%%%%%%%%%%%%%%%%%%%%%%%%%%%%%%%%%%%%%%%%%%%%%%%%%%%%%%%%%%%%%%%%%%%%%%

\noindent
{$\underline{\ttau^2 \mid \mbox{rest}}$}

\noindent
Let $r_{j}(t) = x_j(t) - \mu_{j,\xi_j(t)} - z_j(t)$.  Then $r_{j}(t) \stackrel{iid}{\sim} N(0,\ttau^2)$, and the inverse-Gamma prior on $\ttau^2$ is conjugate, leading to the simple update,
\bdm
\ttau^2 \sim \mbox{IG}\left(A_\tau + \frac{1}{2}\sum_{j=1}^J T_j \;,\;
B_\tau + \frac{1}{2}\sum_{j=1}^J \sum_{t=1}^{T_j}r_j(t)^2\right).
\edm
\\[-.1in]

%%%%%%%%%%%%%%%%%%%%%%%%%%%%%%%%%%%%%%%%%%%%%%%%%%%%%%%%%%%%%%%%%%%%%%%%%%%%%%

\noindent
{\underline{MH update for $\talpha_\lambda$}}

\noindent
As mentioned in the main paper, $\talpha_\lambda$ was updated via a MH random walk proposals.  Again, the random walk was conducted on the log scale, i.e., $\log(\talpha_\lambda^*) = \log(\talpha_\lambda+\epsilon)$ for a deviate $\epsilon \sim N(0,s^2)$. A tuning parameter $s^2=0.01$ was used to achieve an acceptance rate of 40\%.  Let the density of the proposal, given the current value of $\talpha_\lambda$ be denoted $d({\talpha_\lambda}^* \mid {\talpha_\lambda})$.  The only portion of the full model likelihood that differs between the current value and the proposal is
\bdm
\cL(\blambda; \talpha_\lambda, \tbeta_\lambda) = \prod_{j=1}^J \prod_{k=1}^K \mbox{Beta}(\lambda_{j,k}; \talpha_\lambda, \tbeta_\lambda).
\edm
The MH ratio is then
\bdm
MH = \frac{\cL(\blambda; \talpha_\lambda^*, \tbeta_\lambda) \pi(\talpha_\lambda^*) d(\talpha_\lambda \mid \talpha_\lambda^*)}
{\cL(\blambda; \talpha_\lambda, \tbeta_\lambda) \pi(\talpha_\lambda) d(\talpha_\lambda^* \mid \talpha_\lambda)}
\edm
\\[-.1in]

%%%%%%%%%%%%%%%%%%%%%%%%%%%%%%%%%%%%%%%%%%%%%%%%%%%%%%%%%%%%%%%%%%%%%%%%%%%%%%

\noindent
{\underline{MH update for $\tbeta_\lambda$}}

\noindent
The update for $\tbeta_\lambda$ was conducted in a completely analogous manner to that for $\talpha_\lambda$ above.
\\[-.1in]

\end{appendix}

\end{document}